\documentclass[a4paper]{aa}
\usepackage{graphics,epsfig,txfonts}
\usepackage[utf8]{inputenc}
\usepackage[section]{placeins}
\usepackage{amsmath}
\usepackage{color}
\usepackage{xcolor}
\usepackage{natbib}
\usepackage[export]{adjustbox}
\bibpunct{(}{)}{;}{a}{}{,}

\def \MSUN{{\rm M}_{\odot}}
\def\vmeas {\ifmmode{\mathrm{v}} \else {$\mathrm{v}$}\fi} 
\def\vsys {\ifmmode{V_\mathrm{sys}} \else {$V_\mathrm{sys}$}\fi} 
\def\vrot {\ifmmode{V_\mathrm{rot}} \else {$V_\mathrm{rot}$}\fi} 
\def\PAkin {\ifmmode{PA_\mathrm{{kin}}} \else {$PA_\mathrm{{kin}}$}\fi}  
\def\PAphot {\ifmmode{PA_\mathrm{{phot}}} \else {$PA_\mathrm{{phot}}$}\fi}  
\def\re {\ifmmode{R_\mathrm{{e}}} \else {$R_\mathrm{{e}}$}\fi}

\def \rsoft {\ifmmode{r_\mathrm{{soft}}} \else {$r_\mathrm{{soft}}$}\fi} 
\title{The stellar halos of ETGs in the IllustrisTNG simulations: the photometric and kinematic diversity of galaxies at large radii}
\author{C. Pulsoni\inst{1,2}
     \and {O. Gerhard\inst{1}}
     \and {M. Arnaboldi\inst{3}}
     \and {A. Pillepich\inst{4}}
     \and {D. Nelson\inst{5}}
     \and {L. Hernquist\inst{6}}
     \and {V. Springel\inst{5}}
     }
     
\titlerunning{The stellar halos of ETGs in the IllustrisTNG simulations}
\authorrunning{C. Pulsoni et al.}

\institute{Max-Planck-Institut f\"ur extraterrestrische Physik, Giessenbachstra{\ss}e, 85748 Garching, Germany
	   \and Excellence Cluster Universe, Boltzmannstra{\ss}e 2, 85748 Garching, Germany 
	   \and European Southern Observatory, Karl-Schwarzschild-Stra{\ss}e 2, 85748 Garching, Germany
	   \and Max-Planck-Institut für Astronomie, K\"onigstuhl 17, 69117 Heidelberg, Germany
	   \and Max-Planck-Institut f\"ur Astrophysik, Karl-Schwarzschild-Str. 1, 85748 Garching, Germany
	   \and Harvard-Smithsonian Center for Astrophysics, 60 Garden Street, Cambridge, MA 02138, USA}

\date{17 June 2020}

\abstract{
    Early-type galaxies (ETGs) are found to follow a wide variety of merger and accretion histories in cosmological simulations.
}{
    We characterize the photometric and kinematic properties of simulated ETG stellar halos, and compare them to observations.
}{
    We select a sample of 1114 ETGs in the 
    TNG100 simulation, and 80 in the higher-resolution TNG50. These
    ETGs span a stellar mass range of $10^{10.3}-10^{12} \MSUN{}$ and are selected within the range of $g-r$ colour and $\lambda$-ellipticity diagram populated by observed ETGs. We determine photometric parameters, intrinsic shapes, and kinematic
    observables in their extended stellar halos. We compare the results
    with central IFU kinematics and ePN.S planetary nebula velocity fields at large radii, study the variation in kinematics from center to halo, and connect it to a change in the intrinsic shape of the galaxies.
}{
    We find that the simulated galaxy sample reproduces the diversity of kinematic properties observed in ETG halos. Simulated fast rotators (FRs) divide almost evenly in one third having flat $\lambda$ profiles and high halo rotational support, a third with gently decreasing profiles, and another third with low halo rotation. However, the peak of rotation occurs at larger $R$ than in observed ETG samples. Slow rotators (SRs) tend to have increased rotation in the outskirts, with half of them exceeding $\lambda=0.2$. For $M_{*}>10^{11.5}M_{\odot}$ halo rotation is unimportant. A similar variety of properties is found for the stellar halo intrinsic shapes. Rotational support and shape are deeply related: the kinematic transition to lower rotational support is accompanied by a change towards rounder intrinsic shape. Triaxiality in the halos of FRs increases outwards and with stellar mass. Simulated SRs have relatively constant triaxiality profiles.
}{
    Simulated stellar halos show a large variety of structural properties, with quantitative but no clear qualitative differences between FRs and SRs. At the same stellar mass, stellar halo properties show a more gradual transition and significant overlap between the two families, despite the clear bimodality in the central regions. This is in agreement with observations of extended photometry and kinematics.
}


\begin{document}

\maketitle

\section{Introduction}

The family of early type galaxies (ETGs) encompasses galaxies that have typically ceased their star formation at early times, with red colors and small amounts of cold gas and dust today, and that mainly consist of elliptical and lenticular galaxies \citep[][]{1994ARA&A..32..115R, 2003MNRAS.341...54K, 2009ARA&A..47..159B}.
Ellipticals essentially divide into two classes with distinct physical properties \citep[e.g.][and references therein]{2009ApJS..182..216K}: those with low to intermediate masses and coreless luminosity profiles, that rotate rapidly, are relatively isotropic and oblate-spheroidal, and have high ellipticities and disky-distorted isophotes; and those which are frequently among the most massive galaxies, with cored profiles, mostly non-rotating, anisotropic and triaxial, relatively rounder than than coreless systems, and with boxy-distorted isophotes. Thus the dichotomy in the light distributions of the ellipticals roughly corresponds to different kinematic properties, with coreless disky objects being rotationally supported, and cored boxy galaxies having low rotation \citep{1987MitAG..70..226B}.
With the advent of integral field spectroscopy (IFS) the classification of elliptical galaxies has shifted to a kinematics-based division between fast rotators (FR) and slow rotators (SR) \citep{2011MNRAS.414..888E, 2018MNRAS.477.4711G}. In particular low mass, coreless, FR ellipticals share similar properties with lenticular galaxies, which are are also included in the FR family, while massive cored ellipticals are typically SRs.

The formation of massive ETGs is believed to have occurred in two phases \citep[e.g.][]{2010ApJ...725.2312O}. In an initial assembly stage, gas collapses in dark matter halos and forms stars in a brief intense burst which is quickly quenched \citep[e.g.][]{2005ApJ...621..673T, 2014ApJ...780...33C, 2010ApJ...721..193P}.
Present-day simulations agree in that the progenitors of FR and SR at these high redshifts are indistinguishable \citep[][with Illustris, Eagle, and Magneticum, respectively]{2017MNRAS.468.3883P, 2017MNRAS.464.3850L, 2018MNRAS.480.4636S}. At $z\lesssim1$ the accretion-dominated phase overtakes, whereby ETGs grow efficiently in size through a series of merger episodes, mainly dry minor mergers \citep{2009ApJ...699L.178N, 2012ApJ...754..115J}, which enrich the galaxies with accreted (ex-situ) stars.
The $\Lambda$CDM cosmology predicts that structures form hierarchically, in which more massive systems form through the accretion of less massive objects. This means that more massive galaxies can have accreted fractions larger than 80\%, while lower mass galaxies are mostly made of in-situ stars, and the accreted components are mainly deposited in the outskirts \citep{2016MNRAS.458.2371R, 2018MNRAS.475..648P}. 
The slow/fast rotators (i.e. the core/coreless) classes result from different formation pathways characterized by different numbers of mergers, merger mass ratio, timing, and gas fractions \citep[][see also the discussion in Kormendy et al. 2009]{2014MNRAS.444.3357N, 2017MNRAS.468.3883P}, although the details still depend on the star formation and AGN feedback models adopted by the numerical models \citep{2017ARA&A..55...59N}. In general, the result of a formation history dominated by gas dissipation is most likely a coreless FR, while dry major mergers often result in SRs. 

The two-phase formation scenario is supported both by observations of compact red nuggets at $z\sim2$, a factor of 2-4 smaller than present day ellipticals \citep{2005ApJ...626..680D,2007MNRAS.382..109T, 2008ApJ...677L...5V}, and by evidence for a subsequent rapid size growth with little or no star formation \citep[e.g.][]{2010ApJ...709.1018V,2011ApJ...739L..44D, 2014ApJ...788...28V,2017MNRAS.466.4888B}. The merger driven size growth is supported by the observed rate of mergers from pair counts and identified interacting galaxies \citep{2008ApJS..175..356H,2010ApJ...719..844R}, as well as the observed tidal debris from recent accretion events in the halos of many galaxies \citep[e.g.][]{1983ApJ...274..534M, 2010ApJ...715..972J, 2015A&A...579L...3L, 2017ApJ...839...21I, 2019A&A...632A.122M}.

A consequence of the two-phase formation is that ETGs are layered structures in which the central regions are the remnants of the stars formed in-situ, while the external stellar halos are principally made of accreted material \citep{2005ApJ...635..931B, 2010MNRAS.406..744C}, even though the details strongly depend on stellar mass \citep{2018MNRAS.475..648P}. Because of the different nature of the stellar halos, galaxies are expected to show significant variation of physical properties from central regions to large radii, such as shapes of the light profiles \citep{2013ApJ...768L..28H, 2014MNRAS.443.1433D,2017A&A...603A..38S}, stellar populations \citep{2014MNRAS.442.1003P, 2020MNRAS.491.3562Z}, and kinematics \citep{2009MNRAS.394.1249C, 2012ApJS..203...17R,2014ApJ...791...80A,2016MNRAS.457..147F}.

Kinematic measurements in the outer halos of ETGs require alternative kinematic tracers to overcome the limitations from the faint surface brightness in these regions, such as planetary nebulae (PNe)  \citep[e.g., the ePN.S survey,][see Sect. \ref{sec:observed_quantities}]{2017IAUS..323..279A}, or globular clusters \citep[e.g., the SLUGGS survey,][]{2014ApJ...796...52B}.
Recently \citet{2018A&A...618A..94P} found evidence from the ePN.S survey for a kinematic transition between the central regions and the outskirts of ETGs. Despite the FR/SR dichotomy of their central regions, these ETG halos display a variety of kinematic behaviors. A considerable fraction of the ePN.S FRs show reduced rotational support at large radii, which has been interpreted as the fading of a rotating, disk-like component into a more dispersion dominated spheroid; almost half of the FR sample shows kinematic twists or misalignments at large radii, indicating a variation of their intrinsic shapes, from oblate at the center to triaxial in the halo. SRs instead have increased rotational support at large radii. While a smaller group of FRs stands out for having particularly high $V/\sigma$ ratio in the halo, most of the ePN.S FRs and SRs have similar $V/\sigma$ ratio in the halo regions. These results suggest the idea that at large radii the dynamical structure of these galaxies could be much more similar than in their high-density centers: if halos are mainly formed from accreted material, their common origin would explain their similarities. The radii of the observed kinematic transitions to the halo and their dependence on the galaxies' stellar mass seem to support such an interpretation. 

Up to date only a few studies of the kinematic properties of stellar halos in simulations are available in the literature. \citet{2014MNRAS.438.2701W} analysed the kinematics of 42 cosmological zoom simulations of galaxies and found a variety of $V/\sigma$ profile shapes (rising, flat, or with a maximum), in agreement with observations. However these early simulations did not reproduce the whole spectrum of properties of observed FRs, especially the fast rotating and extended disks \citep[][]{2011MNRAS.414..888E,2018A&A...618A..94P}. Recently, \citet{2020MNRAS.493.3778S} using the Magneticum Pathfinder simulations showed that these simulations reproduce the observed kinematic properties of galaxies more closely, and that extended kinematics is a valuable tool for gaining insight into galaxy accretion histories. They also found that the kinematic transition radius is a good estimator of radius of the transition between in-situ and ex-situ dominated regions for a subset of galaxies with decreasing $V/\sigma$ profiles, especially those that did not undergo major mergers in their evolution.

The goal of this paper is to better understand the structural changes between the centers and stellar halos of ETGs with a large and well-resolved sample of simulated galaxies. 
We study the stellar halo structure, i.e., the rotational support and intrinsic shapes of the simulated galaxies, we compare the results with observations, and we investigate how the radial variations in rotational support relate to changes in the halo shapes.
We use the IllustrisTNG simulations \citep{2018MNRAS.475..676S,  2018MNRAS.475..648P,2018MNRAS.477.1206N, 2018MNRAS.480.5113M, 2018MNRAS.475..624N, 2019ComAC...6....2N}, a suite of magnetohydrodynamical simulations that models the formation and evolution of galaxies within the $\Lambda$CDM paradigm. It builds and improves upon the Illustris simulation \citep{2014MNRAS.445..175G, 2014MNRAS.444.1518V}, using a refined galaxy formation model.
For this work we consider two cosmological volumes with side lengths $\sim100$ Mpc and $\sim50$ Mpc, which are referred to as TNG100 and TNG50. TNG50 is the highest resolution realization of the IllustrisTNG project \citep{2019MNRAS.490.3196P,2019MNRAS.490.3234N} with particle resolution more than 15 times better than TNG100.

The paper is organized as follows. For comparing the TNG galaxies properties with observations, we first summarize in Section \ref{sec:observed_quantities} how different ETG surveys select their samples, and how physical quantities are measured. Section \ref{sec:method} then describes and illustrates our methods to derive photometric and kinematic measurements for the simulated galaxies. After selecting the sample of ETGs from the TNG100 and TNG50 simulations (Section \ref{sec:SelectionSample0}), we proceed to show the photometric results in Section \ref{sec:Photometry} and the kinematic results in Section \ref{sec:Kinematics}. Section \ref{sec:AM_shape} relates the variation in the kinematic properties from central regions to halos to the parallel changes in the intrinsic structure of galaxies. In a companion paper we will explore the dependence of these properties on the accretion history of galaxies. Finally Section \ref{sec:summary_and_conclusions} summarizes our conclusions.
\section{The IllustrisTNG simulations}
\label{sec:simulations}

\begin{table}
\caption{ Table of physical and numerical parameters for TNG50 and TNG100. These are the volume of the box, the initial number of particles (gas cells and dark matter particles), the target baryon mass, the dark matter particle mass, and the $z=0$ Plummer equivalent gravitational softening length for the collisionless component.}             
\label{tab:SimulationsParams}  
\centering    
\begin{tabular}{ c c c c c c c }
\hline\hline      
 Run     & Volume & N$_{\rm part}$ & m$_{\rm baryons}$ & m$_{\rm DM}$ & $\rsoft{}$\\
 name  & [Mpc$^3$]     &              & [$10^5 M_\odot$]       &  [$10^5 M_\odot$] &  [pc] \\
 \hline
 TNG50   &  $51.7^3$         & $2\times2160^3$      & $0.85$   &$4.5$ & 288    \\
 TNG100     & $110.7^3$         & $2\times1820^3$ & $14$   & $75$  & 738 \\
\hline  
\end{tabular}
\end{table}

The IllustrisTNG simulations are a new generation of cosmological magnetohydrodynamical simulations using the moving mesh code \textsc{arepo} \citep{2010MNRAS.401..791S}. Compared to the previous Illustris simulations, they include improvements in the models for chemical enrichment, stellar and black hole feedback, and introduce new physics such as the growth and amplification of seed magnetic fields.

The baryonic physics model contains a new implementation of black hole feedback \citep{2017MNRAS.465.3291W}, as well as updates to the galactic wind feedback, stellar evolution and gas chemical enrichment models \citep{2018MNRAS.473.4077P}. These modifications, in particular those for the two feedback mechanisms, were required to alleviate some of the tensions between Illustris and observations, such as the large galaxy stellar masses below the knee of the galaxy stellar mass function and the gas fractions within group-mass halos. They in turn also improve on the too large stellar sizes of galaxies and the lack of a strong galaxy color bimodality at intermediate and high galaxy masses in Illustris \citep{2015A&C....13...12N}.

The IllustrisTNG fiducial model was chosen by assessing the outcome of many different models against the original Illustris by using additional observables, specifically the halo gas mass fraction and the galaxy half-mass radii, with respect to those used to calibrate the Illustris model against observational findings, i.e. the star formation rate density as a function of $z$, the galaxy stellar mass function at $z = 0$,  the $z=0$ black hole mass versus halo mass relation, and the $z = 0$ stellar-to-halo mass relation. 

The new AGN feedback model is responsible for the quenching of galaxies in massive halos and for the production of red and passive galaxies at late times,  alleviating the discrepancies with observational data at the massive end of the halo mass function \citep{2017MNRAS.465.3291W, 2018MNRAS.475..624N, 2019MNRAS.485.4817D}. The faster and more effective winds in TNG reduce the star formation at all masses and all times, resulting in a suppressed $z=0$ galaxy stellar mass function for $M_{*}\lesssim10^{10}\MSUN$, and smaller galaxy sizes \citep{2018MNRAS.473.4077P}. Overall the TNG model has been demonstrated to agree satisfactorily with many observational constraints \citep[e.g.][]{2018MNRAS.474.3976G, 2018MNRAS.475..624N} and to return a reasonable mix of morphological galaxy types \citep{2019MNRAS.483.4140R}.

In this study we consider two simulation runs, TNG100 and TNG50, which are the two highest resolution realizations of the IllustrisTNG intermediate and small cosmological volumes. TNG100 has a volume and resolution comparable with Illustris, while TNG50 reaches resolutions typical of zoom-in simulations. Table~\ref{tab:SimulationsParams} summarizes and compares the characteristic parameters of the two simulations. 

The TNG model is calibrated at the resolution of TNG100 and all the TNG runs adopt identical galaxy formation models with parameters that are independent of particle mass and spatial resolution \citep["strong resolution convergence" according to][]{2015MNRAS.446..521S}. This imposition results in some of the properties of the simulated galaxies being resolution dependent.
As discussed by \cite{2018MNRAS.473.4077P}, this can be primarily explained by the fact that better resolution allows the sampling of higher gas densities, hence more gas mass is eligible for star formation and the star formation rate accelerates. 
This means that, for example, at progressively better resolution, galaxies tend to have increased stellar masses at fixed halo mass and smaller sizes at fixed stellar mass (see also \citealt{2019MNRAS.490.3196P} for a quantification of these effects).


\section{Observed parameters of ETGs}
\label{sec:observed_quantities}

In this paper we compare the kinematic results for the central regions of the simulated TNG galaxies with IFS measurements from the surveys Atlas3D \citep{2011MNRAS.413..813C}, MANGA  \citep{2015ApJ...798....7B}, SAMI \citep{2012MNRAS.421..872C} and MASSIVE \citep{2014ApJ...795..158M}. 

Kinematics measurement at large radii are notably difficult to obtain for ETGs, and therefore discrete kinematic tracers such as planetary nebulae (PNe) and globular clusters (GCs) are typically used to overcome the limitations of absorption line spectroscopy, which is restricted to the central 1-2 \re{}. PNe are established probes of the stellar kinematics in ETG halos \citep{1995ApJ...449..592H, 1996ApJ...472..145A, 2001ApJ...563..135M, 2009MNRAS.394.1249C, 2013A&A...549A.115C}, out to very large radii \citep{2015A&A...579A.135L,2018A&A...616A.123H}. Since they are drawn from the main stellar population, their kinematics traces the bulk of the host-galaxy stars, and are directly comparable to integrated light measurements. 
The relation between GCs and the underlying galaxy stellar population is less straightforward \citep{2018MNRAS.479.4760F}. In general GCs do not necessarily follow the surface brightness distribution and kinematics of the stars \citep[e.g.][]{2006ARA&A..44..193B,2013MNRAS.436.1322C, 2014MNRAS.442.2929V}, although there is growing evidence for red, metal-rich GCs to be tracers of the host galaxy properties \citep{2020A&A...637A..26F, 2020MNRAS.495.1321D}. Therefore we here compare the kinematics of the simulated galaxies and their stellar halos at large radii with PN kinematic results from the ePN.S early-type galaxy survey \citep[][Arnaboldi et. al., in prep.]{2017IAUS..323..279A}. 

Below we describe the sample properties for the different surveys, and we give details and sources of the measured quantities used though out this paper.

\emph{Sample properties} - The Atlas3D survey selected ETGs from a volume-limited sampe of galaxies, with distance within 42 Mpc, and sky declination $\delta$ such that  $|\delta-29^\circ|<35^\circ$), brighter than $M_K<-21.5$ mag. From this parent sample ETGs were morphologically selected as all the galaxies without visible spiral structure. This morphological selection is broadly similar to a selection of the red sequence \citep{2011MNRAS.413..813C}. The Atlas3D ETG sample contains 68 Es and 192 S0s.
The SAMI survey \citep{2012MNRAS.421..872C} selected a volume and magnitude limited sample of galaxies in the redshift range $0.004<z<0.095$, covering a broad range in galaxy stellar mass ($M_{*}=10^8-10^{12}\MSUN{}$) and environment (field, group, and clusters). This sample is not morphologically selected, but we use the data from \cite{2017ApJ...835..104V} where the quality cuts and the imposed threshold on the velocity dispersion $\sigma>70$ km/s bias the sample towards the ETGs (82\%). 
The galaxies of the MANGA survey \citep{2015ApJ...798....7B} are selected from the NASA-Sloan Atlas\footnote{http://www.nsatlas.org} (NSA) catalog \citep[which is based on the  Sloan Digital Sky Survey (SDSS) Data Release 8,][]{2011ApJS..193...29A} at low redshift ($0.01<z<0.15$), to follow a flat distribution in stellar mass in the range $M_{*}=10^9-10^{12}$; in this paper we will compare only with MANGA's galaxies classified as ellipticals or lenticulars as in \cite{2018MNRAS.477.4711G}.
The MASSIVE survey \citep{2014ApJ...795..158M} targets all the most massive ETGs ($M_{*}\gtrsim10^{11.5} \MSUN{}$) within a distance of 108 Mpc. 
Finally, the ePN.S sample of ETGs is magnitude limited $M_{K}\lesssim-23$, and includes objects with different structural parameters. This ensures the sample to be a representative group of nearby ETGs. The ePN.S kinematic results \citep{2018A&A...618A..94P} combine PN kinematics in the halos with literature absorption line data for the central regions.

\emph{Colors} - The MANGA galaxies, and most of the Atlas3D and MASSIVE objects have measured $g-r$ colors in the NSA catalog. 
For the SAMI galaxies \cite{2017ApJ...835..104V} report $g-i$ colors, which we convert to $g-r$ using the transformation equation derived in App. \ref{sec:color_relations}. For all of the ePN.S sample, and some of the Atlas3D and MASSIVE galaxies that are not in the NSA catalog, we use $B-V$ colors corrected for galactic extinction from the Hyperleda\footnote{http://www.leda.univ-lyon1.fr} catalog \citep{2014A&A...570A..13M}, and convert to $g-r$ colors using the relations in App. \ref{sec:color_relations}.

\emph{Sizes} - For the Atlas3D sample we use the effective radii ($R_e$) values in Table 3 of \citep{2011MNRAS.413..813C}. 
Those for the MASSIVE galaxies are from \citet[][Table 3]{2014ApJ...795..158M}, where we adopt the NSA measurements, where available, or the 2MASS values corrected using their Equation 4. 
The data for MANGA are from \cite{2018MNRAS.477.4711G}. 
For SAMI we use the data presented in \cite{2017ApJ...835..104V}, and we circularise the effective semi-major axis by using the reported value for the ellipticity. 
The half light radii for the ePN.S galaxies are in Table 2 of \cite{2018A&A...618A..94P}. These are effective semi-major axis distances measured from the most extended photometric profiles available from the literature, extrapolated to very large radii with a S{\'e}rsic fit. The ellipticity assumed is in their table 1. Section \ref{sec:measuring_Re} discusses the systematic effects in comparing observed effective radii and half-mass radii in simulated galaxies.

\emph{Stellar masses} - The IllustrisTNG model assumes a  \cite{2003PASP..115..763C} initial mass function (IMF). The stellar masses for the SAMI survey in \cite{2017ApJ...835..104V} are derived using a color–mass relation, and a Chabrier IMF. For Atlas3D, MASSIVE, and MANGA we use the total absolute K-band luminosity $M_{K}$ from the same tables referenced above, which are derived from the 2MASS extended source catalog \citep[][]{2003AJ....125..525J}, and already corrected for galactic extinction. The luminosities $M_{K}$ are then corrected for missing flux as in \citet{2013ApJ...768...76S}, $M_{K_{corr}} = 1.07 M_{K} +1.53$, and converted to stellar masses with the formula from \cite{2019MNRAS.484..869V}
\begin{equation}
\log_{10} M_{*} = 10.39 - 0.46 (M_{K_{corr}} + 23),
\label{eq:lightKtomass}
\end{equation}
which uses the stellar population model-based mass-to-light ratio from \citet[][]{2013MNRAS.432.1862C}, their $[\log(M/L)_\mathrm{Salp}]$, converted to a Chabrier IMF. The missing flux correction takes into account the over-subtraction of the sky background by the 2MASS data reduction pipeline \citep{2012PASA...29..174S} and the limited $4\re{}$ aperture of the 2MASS measurement.

For the ePN.S sample we derive stellar masses using integrated luminosities from the most extended photometric profiles available in the literature, extrapolated to infinity with a S{\'e}rsic fit \citep[references in][]{2018A&A...618A..94P}. We convert the integrated values to stellar masses by using the non-dereddened relations between colors and mass-to-light ratios for ellipticals and S0 galaxies from \cite{2019A&A...621A.120G}, which assume a Chabrier IMF. 

There are several sources of errors in the stellar mass estimates of observed galaxies. The uncertainty in the magnitudes derived from the 2MASS photometry are typically $\sim0.25$ mag \citep{2013ApJ...768...76S}. The uncertainty in the distances typically translate into an error of $0.1$ mag on the absolute magnitudes but can reach up to $0.5$ mag \citep{2017ApJ...835..104V}. These uncertainties correspond to an error on the stellar mass of typically $\sim0.1$ dex and up to $\sim0.2$ dex. In addition the total luminosity, and hence the total stellar mass, can be underestimated if the photometry is not deep enough to measure the faint surface brightness of the stellar halos, especially in massive galaxies with large S{\'e}rsic indices or described by multiple S{\'e}rsic components. 
Since the stellar masses of the simulated galaxies are evaluated using the total bound stellar mass (Section \ref{sec:SelectionSample}), this may cause a systematic difference between observed and simulated stellar masses at the high mass end; see also Section \ref{sec:measuring_Re}.

\emph{Ellipticities} - For the Atlas3D galaxies we use the ellipticity $\varepsilon$ measurements within 1 \re{} reported in Table B1 of \cite{2011MNRAS.414..888E}. 17 out of 260 Atlas3D objects have obvious bar components: for these cases the ellipticity is measured at larger radii (typically 2.5 - 3\re{}). 
Ellipticities for the SAMI galaxies are from \cite{2017ApJ...835..104V}, and are average ellipticities of the galaxies within 1 \re{}. MANGA's ellipticities from \cite{2018MNRAS.477.4711G} are also measurements within the 1 \re{} isophote, while for the MASSIVE sample \cite{2017MNRAS.471.1428V} uses ellipticities from NSA where available, and from 2MASS otherwise, which are globally fitted values. 
The ellipticity profiles for the ePN.S galaxies are referenced in \cite{2018A&A...618A..94P}. 
The measurement errors on the ellipticities are per se very small (O($10^{-3}$), \citealt{2009ApJS..182..216K}), but the characteristic ellipticities used by different
surveys for the same galaxies can differ within a root-mean-square scatter of $\sim0.05$ (see e.g. \citealt{2017MNRAS.471.1428V} and figure 2 from \citealt{2018MNRAS.477.4711G}).

\emph{Angular momentum parameters $\lambda_e$} - The parameter $\lambda_e$ is derived in the different surveys using different integration areas. While \citet[][Atlas3D]{2011MNRAS.414..888E} and \citet[][MASSIVE]{2017MNRAS.471.1428V} use circular apertures of radius $\re{}$, \citet[][SAMI]{2017ApJ...835..104V}  prefer elliptical apertures with semi-major axis $\re{}$, and \citet[][MANGA]{2018MNRAS.477.4711G}  integrate over the half-light ellipse (an ellipse covering the same area as a circle with radius $\re{}$, i.e. with semi-major axis $\re{}\sqrt{1-\varepsilon}$, where $\varepsilon$ is the ellipticity).

The uncertainties on the measured $\lambda_e$ for the Atlas3D galaxies are generally small, $\Delta\lambda_e\simeq0.01$ \citep{2011MNRAS.414..888E}.  Similar errors apply for the MASSIVE sample, $\Delta\lambda_e\lesssim0.01$ \citep{2017MNRAS.471.1428V}.
SAMI and MANGA instead target objects at larger distances with lower apparent sizes and spatial resolution. For these galaxies the measurement uncertainties are combined with seeing effects, which generally tend to systematically decrease $\lambda_e$. In the SAMI galaxies, for a typical seeing of 2 arcsec, \citet{2017ApJ...835..104V} find that measurement errors ($\Delta\lambda_e \sim 0.01$) and seeing effects cancel out for galaxies with $\lambda_e < 0.2$, while for $\lambda_e>0.2$ seeing is the dominant effect and causes a median decrease in $\lambda_e$ of $0.05$. For the MANGA regular rotators in the cleaned sample, \citet{2018MNRAS.477.4711G} estimate mean $\Delta\lambda_e = [0.005,-0.041]$ and median errors $\Delta\lambda_e =[0.004,-0.027]$. 

\emph{$V/\sigma$ profiles} - The $V/\sigma$ profiles for the Atlas3D and the ePN.S galaxies are derived from the ratio of the rotation velocity $V_\mathrm{rot}$ and the azimuthally averaged velocity dispersion in elliptical radial bins. For the Atlas3D galaxies we apply the procedure described in Sect. \ref{measuring_VelFields_Halo} directly to the velocity fields from \cite{2004MNRAS.352..721E} and \cite{2011MNRAS.413..813C}, 
giving a median error on $V/\sigma$ of the order of $0.03$.

For the ePN.S galaxies the procedure is applied to the PN velocity fields, whereas for the central regions we use the $V_\mathrm{rot}$ and $\sigma$ from kinemetry analysis on IFS data from \citet{2008MNRAS.390...93K, 2011MNRAS.414.2923K, 2016MNRAS.457..147F}, when available. In the other cases we use $V_\mathrm{rot}$ and $\sigma$ from major axis slits (see references in the ePN.S paper). 
For the ePN.S galaxies the measurement uncertainties on the $V/\sigma$ profiles are dominated by the statistical error on the PN velocity fields. The median $\Delta(V/\sigma) = 0.08$. 

\section{Methods: IllustrisTNG photometry and kinematics}\label{sec:method}

In this section we describe the method for measuring photometry and kinematics in the IllustrisTNG galaxies. 
For each simulated galaxy we define a coordinate system $(x,y,z)$ aligned with the axes of the simulation box, and centered at the position of the most bound particle in the galaxy.
Galaxies are observed both edge-on and along a random fixed line-of-sight (LOS) direction. The edge-on projection is obtained by rotating the particles according to the principal axes of the moment of the inertia tensor $I_{ij}$ 
\begin{equation}
    I_{ij} = \frac{\sum_n M_{n} x_{n,i} x_{n,j}}{\sum_n M_{n}},
\label{eq:inertia_tensor}
\end{equation}
where the sum is performed over the 50\% most bound stellar particles; $x_{n,i}$ is their coordinates, $M_n$ their mass.
The random LOS direction is arbitrarily chosen to be the $z$ axis of the simulation box. 
In this work we will indicate with the lowercase letters $x_i$, $v_i$, and $r_i$ the 3D coordinates, velocities and radii, and we reserve capital letters for the corresponding 2D quantities projected on the sky. The coordinate $r$ indicates the intrinsic semi-major axis distance, while $R$ indicates the projected semi-major axis distance.

For any projection, we rotate the galaxies so that the $X$ axis corresponds to the projected major axis, and the $Y$ axis to the projected minor axis. This is done by evaluating the inertia tensor in Eq.\ref{eq:inertia_tensor} using the 2D projected coordinates, and summing over the 50\% most bound particles.
We choose to weight quantities by the mass and not by luminosity, as the former are not affected by uncertainties from stellar population modeling and attenuation effects e.g. from dust. The difference between mass weighted and luminosity weighted quantities, such as in the $K$ band, is generally small for old stellar populations \citep[e.g.][]{2008MNRAS.389.1924F}.
Radial profiles are shown in units of effective radii $\re$, which are evaluated as described in Sect. \ref{sec:measuring_Re}.

\begin{figure}
    \centering
    \includegraphics[width=\linewidth]{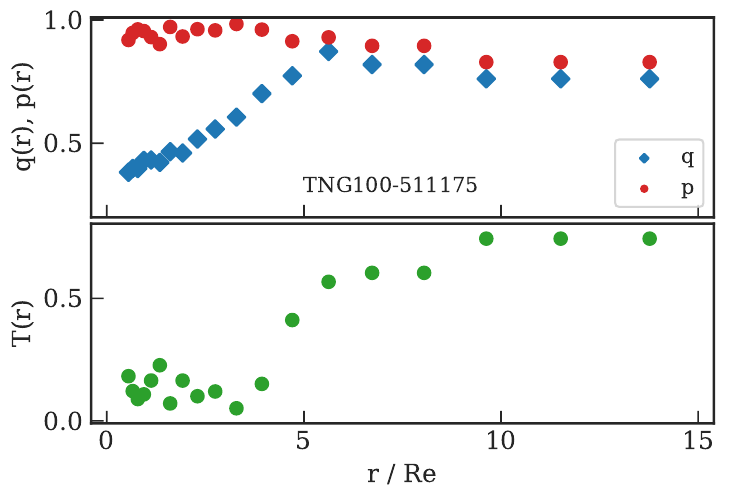}
    \includegraphics[width=\linewidth]{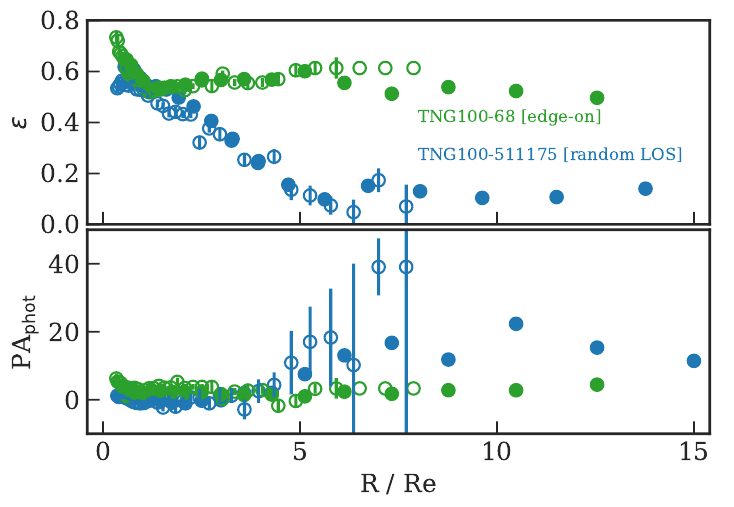}
    \caption{Photometric measurements. \textbf{Top:} intrinsic shape of an example galaxy from TNG100 as a function of the intrinsic major axis distance $r$/\re{}; axis ratios and triaxiality parameter are shown in separate panels. This simulated galaxy is oblate with $q\sim0.4$ and $p\sim0.95$ in the central $3\re{}$, and becomes near-prolate ($T>0.7$) in the outer halo. In this galaxy $9\rsoft{}$ correspond to $0.48\re{}$.
    \textbf{Bottom:} ellipticity and photometric position angle profiles for two example galaxies from TNG100 as a function of the projected major axis distance $R$/\re{}. 
    The quantities derived from the inertia tensor are shown with solid symbols, those from the mock images with open symbols.}
    \label{fig:photometry_example}
\end{figure}

\subsection{Intrinsic shapes}
\label{sec:measuring_IntrinsicShape}

The three-dimensional intrinsic shapes of the galaxies are evaluated by diagonalizing the inertia tensor $I_{ij}$ in Eq. \eqref{eq:inertia_tensor},
summed over stellar particles enclosed in elliptical shells. This definition of $I_{i,j}$ without any weight factors is shown by \citet{2011ApJS..197...30Z} to be the least biased method for measuring the local intrinsic shape of a distribution of particles, and we refer to their work for a detailed description of the procedure. 

In brief, the galaxies are divided in spherical shells of radii $r$ and $r+\Delta r$. In each shell we calculate the tensor $I_{i,j}$: the square root of the ratio of its eigenvalues give the axis ratios $p$ and $q$ (with $p\geq q$) of the principal axes, the eigenvectors their directions. The spherical shell is subsequently deformed to a homeoid of semi-axes $a=r$, $b = p a$ and $c = q a$. We repeat the procedure iteratively until the homeoid is adjusted to the iso-density surface, and the fractional difference
between two iteration steps in both axis ratios is smaller than $1\%$. The values of $p$ and $q$ as functions of the principal major axis length $r$ give the intrinsic shape profiles of the galaxies. We require a minimum number of 1000 particles in each shell as suggested by \cite{2011ApJS..197...30Z}, which assures small errors from particle statistics, and, at the same time, the possibility of measuring intrinsic shape profiles out to at least 8\re{} for $\sim96\%$ of the selected TNG galaxies. 
The directions of the principal axes of the galaxies as functions of the galactocentric distance $r$ are given by the eigenvectors $\hat{e}_j$ (with $j=a,b,c$) of the inertia tensor. 

We also use the triaxiality parameter 
\begin{equation}
    T(r)=\frac{1 - p(r)^2}{1-q(r)^2}
    \label{eq:triaxiality}
\end{equation} 
to quantify the intrinsic shape. 

In App.\ref{appendix:accuracy_shapes_measurement} we find that shape measurements at $1\re{}$ are affected by the resolution of gravitational forces only for the lowest mass galaxies, for which the absolute error on $p$ and $q$ is $\sim0.1$ at the resolution on TNG100. At  $r\sim9\rsoft{}$, i.e. $r \sim3.5\re{}$ for the lowest mass galaxies, and $r\sim 1.1\re{}$ for $M{*}=10^{11}\MSUN{}$, these resolution effects are negligible, and the error on the shape measurements is then due to particle noise and is $\sim0.02$ in TNG100. 
This uncertainty translates into an error of $\Delta T=0.2$ on the $T$ parameter for typical values of the axis ratios in fast rotator ETGs (i.e. $p=0.9$ and $q=0.5$). As discussed in App.\ref{appendix:accuracy_shapes_measurement}, we consider the triaxiality profiles reliable starting from $r=9\rsoft{}$; at smaller radii, where $\Delta T$ is larger, we quantify shapes using $p$ and $q$ which are better defined. 
These results for TNG100 are summarized in Table~\ref{tab:ResolutionAndShape}. For the TNG50 galaxies we expect similar or lower uncertainties.

\begin{table}
\caption{Absolute uncertainties on the shape measurements in TNG100 galaxies: $\Delta p$, $\Delta_q$, and $\Delta T$ at $2\rsoft$ and $9\rsoft{}$ (first two rows), which correspond to different multiples of $\re{}$ for galaxies of different mass as shown in the last three rows. Triaxiality measurements are considered unreliable for $r<9\rsoft{}$. The radius $r_{\rm out}$ indicates the mean radius of the outermost shell that contains at least 1000 particles for at least 95\% of the galaxies within the indicated stellar mass bins.}             
\label{tab:ResolutionAndShape}  
\centering    
\begin{tabular}{ c | c c c c c c }
\hline\hline      
    & $r\sim2\rsoft{}$ & $r\geq9\rsoft{}$ & \\
 \hline
 $\Delta p,\,\Delta q$  & $\lesssim0.1$  & $\sim 0.02$ &  \\
 $\Delta T$  & $-$    & $\sim 0.2$ &  \\
\hline\hline      
    & $2\rsoft{}$ & $9\rsoft{}$ & $r_{\rm out}$\\
\hline
 log$_{10}M_{*}/\MSUN{}:10.3-10.4$   &  $0.76\re{}$  & $3.5\re{}$ & $\sim8\re$ \\
 log$_{10}M_{*}/\MSUN{}:{10.6}-{10.7}$  & $0.47\re{}$  & $2\re{}$ & $\sim8\re$ \\
 log$_{10}M_{*}/\MSUN{} > {11}$  & $<0.25\re{}$  & $\lesssim 1\re{}$ & $\sim11\re$ \\
\hline  
\end{tabular}
\end{table}

In the paper we will consider halos as near-oblate when $T \leq 0.3$, and near-prolate when $T>0.7$. Halos with intermediate values of $T$ parameter are designated as triaxial.

Figure \ref{fig:photometry_example} (top panel) shows the principal axis ratios $q(r)$ and $p(r)$ as a function of the major axis distance $r$ for one example TNG galaxy, normalized by the \re{} of the edge-on projection. The galaxy shown in the example is close to oblate in the central regions, with $q(1\re{})=0.43$ and $p(1\re{})=0.95$ ($T<0.3$). At large radii the galaxy becomes close to prolate with $q(10\re{})=0.76$, $p(10\re{})=0.83$, and triaxiality parameter $T>0.7$. For the galaxy shown $9\rsoft{} = 0.48\re{}$.

\subsection{Ellipticity and photometric position angle profiles}\label{sec:measuring_2Dphot}

Mass weighted photometry is derived by diagonalizing the 2D inertia tensor (Eq. \eqref{eq:inertia_tensor}) using the projected coordinates for a given LOS. We use an iterative procedure similar to the one described in Sect. \ref{sec:measuring_IntrinsicShape} for the 3D intrinsic shape. The square root of the ratio of the two eigenvalues of $I_{ij}$ gives the projected flattening, hence the projected ellipticity $\varepsilon(R)$; the components of the eigenvectors define the photometric position angle $\PAphot{}(R)$. The zero point of the $\PAphot{}(R)$ is chosen to be the $X$ axis of the galaxies.

As an independent check on the results, we derived $\varepsilon(R)$ and $\PAphot{}(R)$ also from fitting ellipses to mock images of the galaxies, and obtained very similar results.  The bottom panels of Fig.  \ref{fig:photometry_example} shows the $\varepsilon(R)$  and $\PAphot{}(R)$ profiles obtained from the inertia tensor (solid symbols) and from the images (open symbols) for two example galaxies. The galaxy TNG100-511175, shown with blue symbols, is the same as the one shown in the top panels of Fig. \ref{fig:photometry_example}: the increased axis ratio $q(r)$ at $r\geq3\re{}$ is reflected in a decreased projected ellipticity. The example also shows that at low ellipticities the uncertainty on the measured $\PAphot{}(R)$ becomes larger, as is well known. We quantified that our method allows us to measure reliably position angles down to ellipticities $\varepsilon=0.1$, where the error $\Delta\PAphot{}(R)$ from particle noise is $\sim 6^\circ$. Below 0.1 $\Delta\PAphot{}(R)$ increases exponentially when $\varepsilon$ decreases towards 0.

\begin{figure}  
    \centering
    \includegraphics[width=\linewidth]{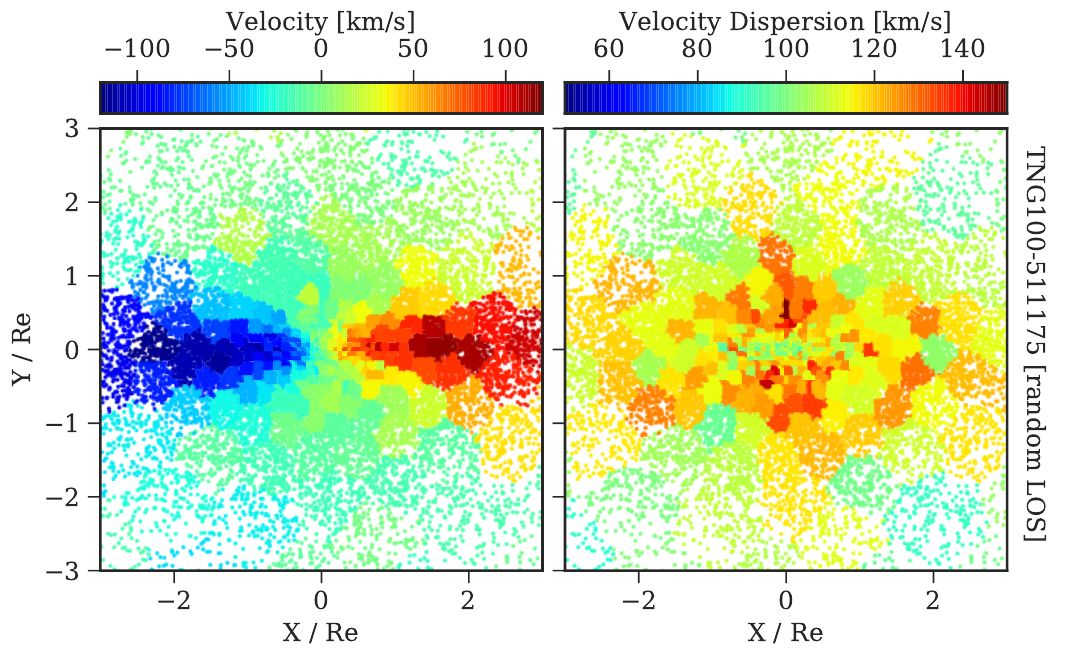}
        \includegraphics[width=\linewidth]{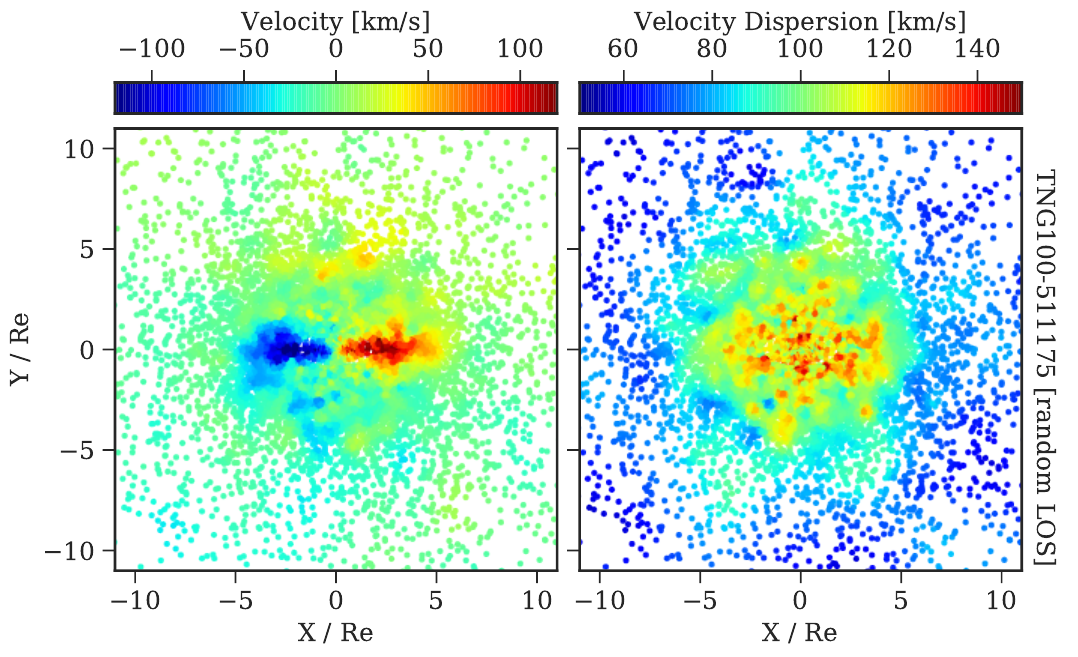}
    \includegraphics[width=0.9\linewidth]{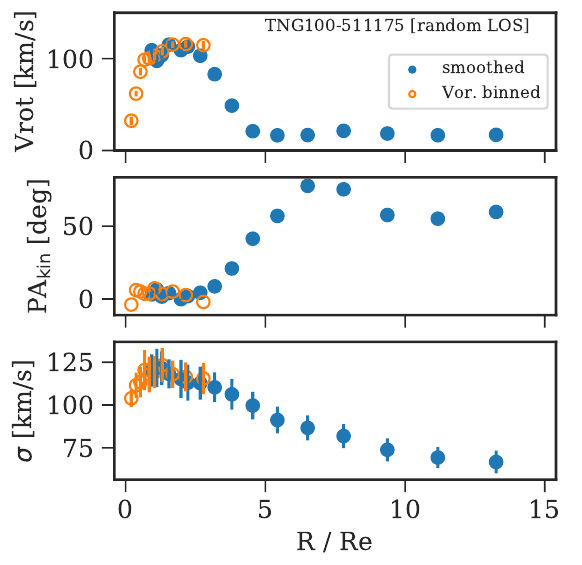}
    \caption{Stellar kinematic measurements. \textbf{Top}: Voronoi binned mean velocity fields for an example TNG100 galaxy, with a central bulge and a relatively massive disk. For this galaxy the random LOS projection almost coincides with the edge-on projection. The projected major axis is aligned with the $X$ axis. The data points show the projected $(X,Y)$ positions of the stellar particles, and are colored according to the mean velocity and velocity dispersion of the corresponding Voronoi bin as shown by the colorbars.
    \textbf{Center}: Smoothed velocity and velocity dispersion fields for the same example galaxy above. The central disk structure is embedded in a spheroidal stellar halo. 
    \textbf{Bottom}: kinematic parameters of the galaxy shown above, derived from the Voronoi binned velocity fields (orange open circles) and from the smoothed velocity fields (blue points).
    }
    \label{fig:velfield_example}
\end{figure}

\subsection{Central kinematics} \label{sec:central_kinematics}

For each TNG galaxy we build projected mean velocity and dispersion fields for two projections (edge-on and random LOS). We use a resolution of $0.2$ kpc, which corresponds to 2 arcsecs for a galaxy observed at 20 Mpc, comparable to present day IFS surveys \citep[e.g.][for MANGA]{2016AJ....152...83L}. The stellar particles are binned on a regular spatial grid centered on the galaxy and 8$\re{}$ wide. 

The binned data are then combined into Voronoi bins as described in \citet{2003MNRAS.342..345C}, so that each bin contains at least 100 stellar particles. In each i-th bin we calculate the projected mean velocity and the mean velocity dispersion as the weighted averages:

\begin{equation}
V_i = \frac{\sum_n M_{n,i} V_{n,i}}{\sum_n M_{n,i}} \; \;\; \; ; \; \; \; \;
\sigma_i  = \sqrt{\frac{\sum_n M_{n,i}\left( V_{n,i} - V_i \right)^2}{\frac{N_i}{N_i-1}\sum_j M_{n,i}}},
\label{eq:meanVS}
\end{equation}
where the index $n$ runs over the particles in the bin, and $N_i$ is the number of particles in the i-th bin.  The top panel of Fig. \ref{fig:velfield_example} shows the result for one example galaxy; the middle and bottom panels show the halo kinematics and the derived kinematic parameters as described in the next section. The example illustrates that in the central regions, where the density of particles is highest, the velocity field is sampled at the highest resolution. At larger radii the Voronoi bins combine the data in progressively larger bins in order to reach the required minimum number of particles.

The systemic velocity of the galaxy is derived by fitting a harmonic expansion as in \citet[][their section 4]{2018A&A...618A..94P} to the central regions (i.e. at $R\leq2\re{}$) of the projected velocity field. The fitted constant term is then subtracted from the velocity fields $V_i$.

From the velocity fields we calculate the angular momentum parameter $\lambda_e$ following the definition of \cite{2011MNRAS.414..888E}
\begin{equation}
  \lambda_e = \frac{\sum_i M_i R_{\mathrm{circ},i}  |V_i|}{\sum_i M_i  R_{\mathrm{circ},i}  \sqrt{V_i^2 + \sigma_i^2}},
\label{eq:lambda}
\end{equation}
where the weighting with the flux is substituted here with a weighting with the mass $M_i$ of each Voronoi bin of index i, $M_i = \sum_{n}M_{n,i}$, and $R_{\mathrm{circ},i}$ is the circular radius of the i-th bin.
The cumulative $\lambda_e$ is derived by summing over all the Voronoi bins contained inside an elliptical aperture of semi-major axis $\re{}$ and flattening given by the ellipticity $\varepsilon(1\re{})$. By comparison, the differential $\lambda(R)$ is summed in elliptical shells. As discussed in App.~\ref{appendix:accuracy_shapes_measurement} the angular momentum parameter is not affected by resolution at $R\gtrsim1\re{}$ for the selected sample of galaxies.

\subsection{Halo kinematics}\label{measuring_VelFields_Halo}

The mean velocity and velocity dispersion fields at large radii are derived using the adaptive smoothing kernel technique \citep{2009MNRAS.394.1249C}, used by \citet{2018A&A...618A..94P} to derive halo velocity fields from the discrete velocities of planetary nebulae in the ePN.S survey. For the simulated galaxies, the discrete velocities of the particles at $R>2\re{}$ are smoothed with a fully adaptive kernel ($A=1$, $B=0$), and their stellar masses are included in the weighting. 

We verified that the kinematic measurements from the adaptively smoothed and the Voronoi binned velocity fields return consistent values in the regions of spatial overlap. The bottom panel of Fig. \ref{fig:velfield_example} shows the rotation velocity $V_\mathrm{rot}$, kinematic position angle $\PAkin{}$, and velocity dispersion $\sigma$ profiles derived from the Voronoi binned velocity fields (in orange), and from the smoothed velocity fields (in blue).  $V_\mathrm{rot}$ and $\PAkin{}$ are derived from fitting a harmonic expansion as in \citet{2018A&A...618A..94P}, and $\sigma$ is azimuthally averaged in elliptical annuli whose flattening follows the ellipticity profile of the galaxies.
The zero point of $\PAkin{}$ is defined to be the $X$ axis of the galaxies, consistently with the zero point of \PAphot{}. 
Error bars on the $\sigma(R)$ profiles are derived from the standard deviation of the $\sigma$ values inside each annulus. The values obtained with the smoothed velocity fields are very well consistent with those from the Voronoi binned velocity fields. 

We also evaluated differential $\lambda$ profiles using Eq. \eqref{eq:lambda}, where the summation is performed over the Voronoi bins and the particles, each weighted by their mass, in elliptical annuli.
We estimated uncertainties on the differential $\lambda(R)$ and on $V/\sigma(R)$ in TNG100 by considering a few kinematically representative galaxies in three stellar mass bins and studied the kinematic parameter distributions derived from 1000 simulations respectively, with particle numbers decreased to the typical numbers at different multiples of $\re{}$. Table~\ref{tab:ResolutionAndKinematics} lists the standard deviation of the distributions for typical numbers of particles at $2\re{}$ and $8\re{}$.

The example galaxy shown in Fig. \ref{fig:velfield_example} has a massive disk ($q\lesssim0.4$, see Fig.  \ref{fig:photometry_example}, top panels) embedded in a spheroidal halo with high T ($T\gtrsim0.7$). The variation in intrinsic shape from near-oblate in the center to strongly triaxial at large radii is accompanied by a modest photometric twist (Fig.  \ref{fig:photometry_example}, bottom panels), and a much larger kinematic twist (Fig.  \ref{fig:velfield_example}) which follows the rotation along the projected minor axis visible in the top panel. At the same radii the rotation velocity $V_\mathrm{rot}$ is observed to drop, together with the local $\lambda$ parameter. 

\begin{table}
\caption{Absolute uncertainties on the kinematic parameters $\lambda(R)$ and $V/\sigma(R)$ for different particle number at the resolution of TNG100.}             
\label{tab:ResolutionAndKinematics}  
\centering    
\begin{tabular}{ c| c c c }
\hline\hline      
log$_{10}\frac{M_{*}}{M_{\odot}}:$ & $10.3-10.5$ & $10.5-11$ & $11-12$\\
 \hline
 $\Delta\lambda(2\re{})$ &  $\sim0.006$  & $\sim0.004$ & $\lesssim0.002$ \\
 $\Delta(V/\sigma)(2\re{})$ & $\sim0.01$ & $\sim0.007$ & $\sim0.003$ \\ 
 \hline
 $\Delta\lambda(8\re{})$   &  $\sim0.02$  & $\sim0.013$ & $\lesssim0.007$ \\
 $\Delta(V/\sigma)(8\re{})$ & $\sim0.03$ & $\sim0.02$ & $\sim0.01$ \\ 
\hline  
\end{tabular}
\end{table}


\section{Selection of the sample of ETGs in the IllustrisTNG simulations}\label{sec:SelectionSample0}

\subsection{Selection in color and mass} \label{sec:SelectionSample}

The purpose of this paper is to study the stellar halos of a volume- and stellar mass-limited sample of simulated ETGs, and compare with observations. 
\citet{2018MNRAS.475..624N} verified that TNG100 reproduces well the $(g-r)$ color of $10^{10}<M_{*}/\MSUN<10^{12.5}$ galaxies at $z=0$, by comparing with the observed distribution from SDSS \citep{2001AJ....122.1861S}. They also showed that redder galaxies have lower star formation rates, gas fractions, gas metallicities, and older stellar populations, and that they correspond to earlier morphological types (their Figure 13).

\begin{figure}[ht] 
    \centering
    \includegraphics[width=\linewidth]{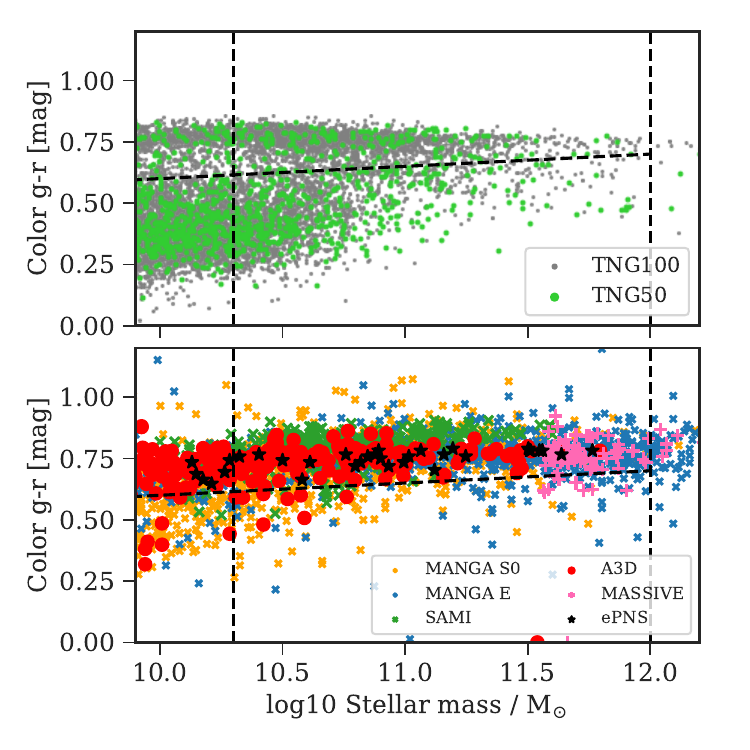}
   
    \caption{Selection of galaxies in color and stellar mass. \textbf{Top:} the $g-r$ color - stellar mass diagram of the simulated galaxies in TNG50 and TNG100. Red sequence galaxies are selected above the tilted dashed line, in the mass range $10^{10.3}<M_{*}/\MSUN<10^{12}$. \textbf{Bottom:} ETGs from recent IFS surveys as indicated. Most of the observed ETGs in this mass range fall in the same red sequence region.}
    \label{fig:SelectionSample}
\end{figure}

\begin{figure} 
    \centering
    \includegraphics[width=\linewidth]{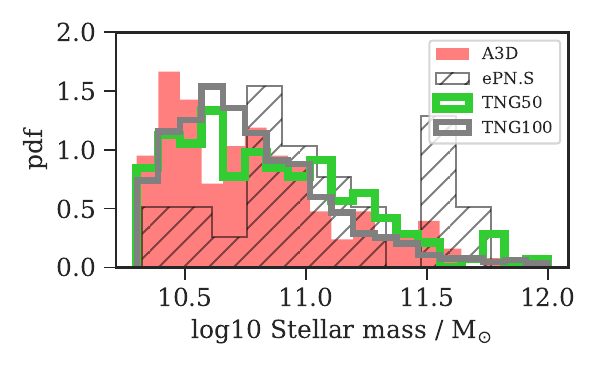}
    \includegraphics[width=\linewidth]{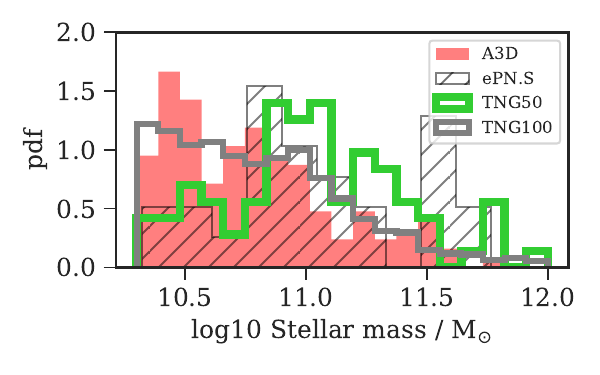}
    \caption{Stellar mass function of the TNG galaxies compared with those of the Atlas3D and ePN.S surveys. 
    \textbf{Top}: stellar mass function of the TNG galaxies selected in color and stellar mass, and whose effective radii are well-resolved, as described in Sect. \ref{sec:SelectionSample}. The stellar mass functions of both TNG50 and TNG100 agree well with Atlas3D.
    \textbf{Bottom}: stellar mass function of the final sample of ETGs, selected in color, stellar mass, and intrinsic shape as described in Sect. \ref{sec:SelectionSample} and Sect. \ref{sec:SelectionSampleII}. The removal of centrally elongated objects mostly changes the low-mass part of the TNG50 mass function, while that of TNG100 is still in good agreement with Atlas3D. }
    \label{fig:MassFunction}
\end{figure}

Thus we extract our sample of ETGs from the TNG50 and TNG100 snapshots at $z = 0$ in the color-stellar mass diagram, isolating galaxies in the red sequence. To obtain a sample of galaxies in the same area occupied by the Atlas3D and the ePN.S samples (see Sect.  \ref{sec:observed_quantities}), we choose 
\begin{equation}
    (g-r) \geq 0.05\log_{10}(M_{*}/\MSUN{}) + 0.1 \mathrm{mag}.
    \label{eq:color_mass_sel}
\end{equation}
For $M_{*}$ we use the total bound stellar mass of the galaxies.
We do not include any dust extinction model in the calculation of the simulated colors in order to avoid the contamination from dust-reddened late type galaxies. Even in this case, this sample of simulated galaxies unavoidably contains some red disks, while in Atlas3D some of the disks have been removed (see Sect. \ref{sec:observed_quantities}). 

We limited the sample stellar mass range to $10^{10.3} \leq M_{*} \leq 10^{12} \MSUN{}$. This choice assures that the TNG100 galaxies are resolved by at least $2\times10^4$ stellar particles. By comparison, the minimum number of stellar particles in the selected TNG50 galaxies is $36\times10^4$. 

In addition we impose that the galaxies' effective radius (see Sect. \ref{sec:measuring_Re}) $\re{} \geq 2\rsoft{}$, to guarantee that the region at $r=\re{}$ is well resolved for all simulated galaxies. For TNG100 $\rsoft{} = 0.74$ kpc at $z=0$, which excludes 38 galaxies at the low mass end (see Fig. \ref{fig:MassSize}). In TNG50 all the galaxies have $\re{}>2\times\rsoft{}$, where $\rsoft{} = 0.288$ kpc.
These criteria select a sample of 2250 galaxies in TNG100 and 168 galaxies in TNG50. 

\begin{figure*}[ht] 
    \centering

    \includegraphics[width=0.90\linewidth]{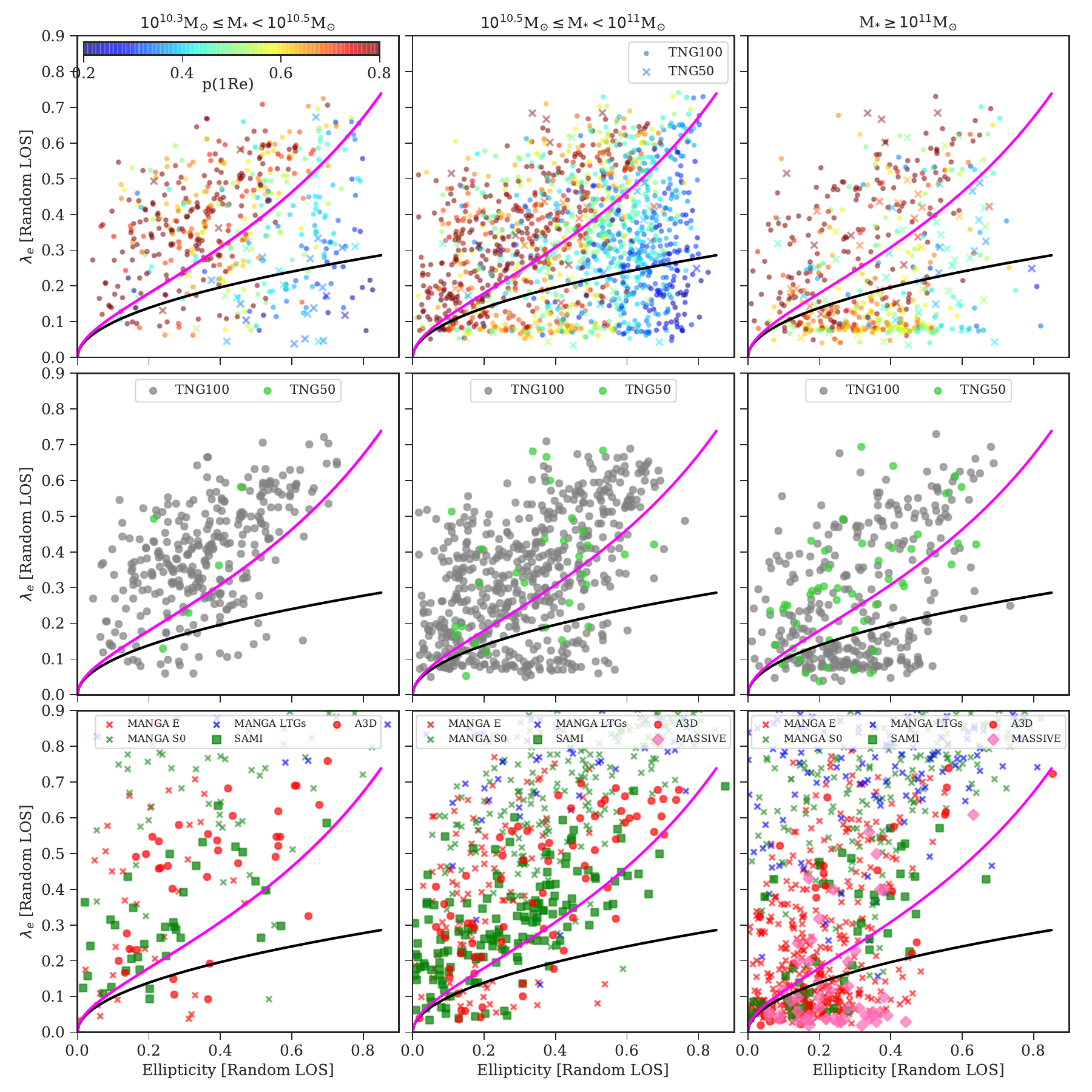}
    \caption{$\lambda_e$-$\varepsilon(1\re{})$ diagrams for TNG and observed galaxy samples, in stellar mass bins. \textbf{Top row}: the sample of TNG50 and TNG100 ETGs, selected by color and mass as in Fig.~\ref{fig:SelectionSample}, observed along a random LOS. The galaxies are color coded according to their intermediate to major axis ratios $p$ at 1Re. \textbf{Middle row}: the $\lambda$-ellipticity diagram for the TNG galaxies after the additional selection $p(1\re{})\geq0.6$. \textbf{Bottom row}: $\lambda$-ellipticity plots for observed ETG samples as shown in the legend, including also late-type galaxies from the MANGA survey for comparison. Their $\lambda_e$ and $\varepsilon(1\re{})$ are as described in Sect.  \ref{sec:observed_quantities}.
    The solid black line is the threshold separating fast and slow rotators as defined in \cite{2011MNRAS.414..888E}: $\lambda_e = 0.31\sqrt{\varepsilon}$. 
    The magenta line shows the semi-empirical prediction for edge-on axisymmetric rotators with anisotropy parameter $\delta=0.70\varepsilon_\mathrm{intr}$ from \cite{2007MNRAS.379..418C}.
    After the additional removal of centrally elongated systems, the colour and mass selected TNG ETG samples give a good representation of the observed ETG samples in the $\lambda_e$-$\varepsilon(1\re{})$-plane, with the exception of the high-$\lambda_e$ ($>0.7$) S0 disks specific to the MANGA survey.
    }
    \label{fig:lambda_ell_center}
\end{figure*}

Figure \ref{fig:SelectionSample} shows the color-stellar mass diagram for the simulated galaxies from TNG100 and TNG50, and for observed galaxies from several IFS surveys. Our selection criteria are highlighted with dashed lines. Most of the observed ETGs, including the SAMI galaxies and the MANGA ellipticals and lenticulars are in the selected region of the diagram. 

The histograms in the top panel of Fig.  \ref{fig:MassFunction} show the stellar mass functions for the color-mass-selected samples. The bottom panel instead shows the stellar mass functions of the final samples as defined by adding constraints from the lambda-ellipticity diagram in Sect. \ref{sec:SelectionSampleII}. The red and hatched histograms show the Atlas3D and ePN.S samples, respectively. Here we consider the Atlas3D sample properties to validate our selection criteria, as this survey is especially targeted to study a volume-limited sample of ETGs. The ePN.S sample, which will be used to compare with properties at large radii, is also shown, and it contains on average higher mass galaxies.
Both TNG50 and TNG100 are in reasonable agreement with Atlas3D. We remark here that a more generous color selection including bluer galaxies would produce a too large number of high ellipticity galaxies especially in TNG50. 

In the following, whenever we compare simulated and observed galaxy samples, we will apply  to the observed galaxies the same color and stellar mass selection criteria that we used for the TNG sample.

\subsection{Selection of ETGs in the $\lambda$-ellipticity diagram: fast and slow rotators} \label{sec:SelectionSampleII}

Figure \ref{fig:lambda_ell_center} shows the $\lambda_e$-$\varepsilon(1\re{})$ diagram for the simulated ETGs in three stellar mass bins, and compares with observed ETG samples. The top row features the diagram for the TNG50 (crosses) and TNG100 (circles) galaxies selected as described in Sect. \ref{sec:SelectionSample}, and projected along a random LOS. The middle row shows again the TNG50 and TNG100 galaxies after the additional selection discussed in this section. The bottom row shows the similar diagram for the observed ETG samples (selected in various ways as described in Sect. \ref{sec:observed_quantities}), in the same color and stellar mass region as defined in Sect. \ref{sec:SelectionSample}. Here we also include for comparison the spiral and irregular galaxies from the MANGA sample (marked as LTGs).

We observe that a significant fraction of the TNG galaxies shown in the top row populate a region to the right of the $\lambda_e$-$\varepsilon(1\re{})$ diagram where there are no observed counterparts, i.e. below the magenta line and with $\varepsilon(1\re{})>0.5$. By color coding the galaxies according to their intrinsic axis ratios at $r\sim1\re{}$, we find that these galaxies have elongated, triaxial shapes. 
These systems occur at all values of $\lambda_e$, i.e. some rotate as rapidly as the MANGA disk galaxies, but others do not show any rotation (Fig. \ref{fig:velocity_fields_elongated}). 

It is possible that some of the rapidly rotating elongated systems are barred galaxies.
\citet{2020MNRAS.491.2547R} showed that within a dynamically selected sample of disk galaxies the TNG100 simulation produces barred systems in fractions consistent with observational results. The majority of these systems, all characterized (per definition) by high rotation, are quenched and hence will overlap with the colour range of our sample of red galaxies.
Some barred galaxies are also expected to be present among the observed ETG samples. For example, in the Atlas3D sample  7\% of the galaxies show a clear bar component. For these objects the $\varepsilon$-values shown in Fig.  
\ref{fig:lambda_ell_center} were measured at larger radii, to avoid the influence of the bar on the estimate of $\varepsilon$ \citep{2011MNRAS.414..888E}. However, if their actual $\varepsilon(1\re{})$ values were used and placed these objects in the region of the $\lambda_e$-$\varepsilon(1\re{})$ populated by the centrally elongated (at $r\sim1\re{}$) TNG galaxies, their fraction would not be large enough to explain the abundance of simulated galaxies in the same region, and none of these have $\lambda_e<0.2$. Therefore the presence of a large fraction of centrally elongated galaxies with high ellipticity and intermediate to low $\lambda_e$ in the TNG sample cannot be explained as a simple sample selection bias (note also that resolution effects on the intrinsic shapes at $1\re{}$ are at most of the order of 0.1, for the low mass galaxies, see App.\ref{appendix:accuracy_shapes_measurement}). 

In App.\ref{appendix:elongated_galaxies} we discuss the properties of these galaxies further, and suggest that they are likely a class of galaxies that are produced by the simulation but are not present in nature.  These galaxies occupy a particular mass range that depends on resolution and they are the reddest systems for their mass. We found no similar concentration of elongated systems among the red galaxies in the Illustris simulation, and the $\lambda-\varepsilon$ diagrams for simulated galaxies in Magneticum \citep{2018MNRAS.480.4636S} and EAGLE \citep{2020MNRAS.494.5652W} do not contain many objects with large ellipticities and intermediate to low $\lambda_e$. This indicates that the new galaxy formation model in TNG is involved in the occurrence of these centrally elongated galaxies. The elongated components typically extend up to 3 $\re{}$ and are embedded in near-oblate spheroids with a wide range of flattening $q$, with lower median value in TNG50 ($q\sim0.3$) than in TNG100 ($q\sim0.45$), indicating a relation to disk building and bar instability. However, some of these do not contain a disk component (Fig.\ref{fig:velocity_fields_elongated}), and they populate a wide range of rotation ($\lambda_e$) approximately uniformly all the way from edge-on $\lambda_e=0.7$ to no rotation (Fig.\ref{fig:lambda_ell_edgeon}). Therefore we suggest that the centrally elongated galaxies in TNG may be systems that were in the process of forming a disk, whose evolution has been interrupted or derailed by rapid dynamical instability, star formation, and feedback in the simulations, in the particular mass range in which they occur.

For these reasons we exclude the centrally elongated objects from our sample of galaxies. We do this by performing a selection in intrinsic shape, and reject all galaxies with intermediate to major axis ratio $p < 0.6$ at $r\sim1R_e$. This choice is motivated by the fact that the intrinsic shape distribution of real galaxies is known \citep{2014MNRAS.444.3340W, 2017MNRAS.472..966F, 2018ApJ...863L..19L, 2018MNRAS.479.2810E} although with large uncertainties \citep{2019MNRAS.487.2354B}, and galaxies with $p < 0.6$ are rare, even among the the slow rotators. 
By applying this selection criterion we obtain our final sample of simulated ETG galaxies, 1114 objects in TNG100 and 80 in TNG50.

\begin{figure}   
    \centering
    \includegraphics[width=\linewidth]{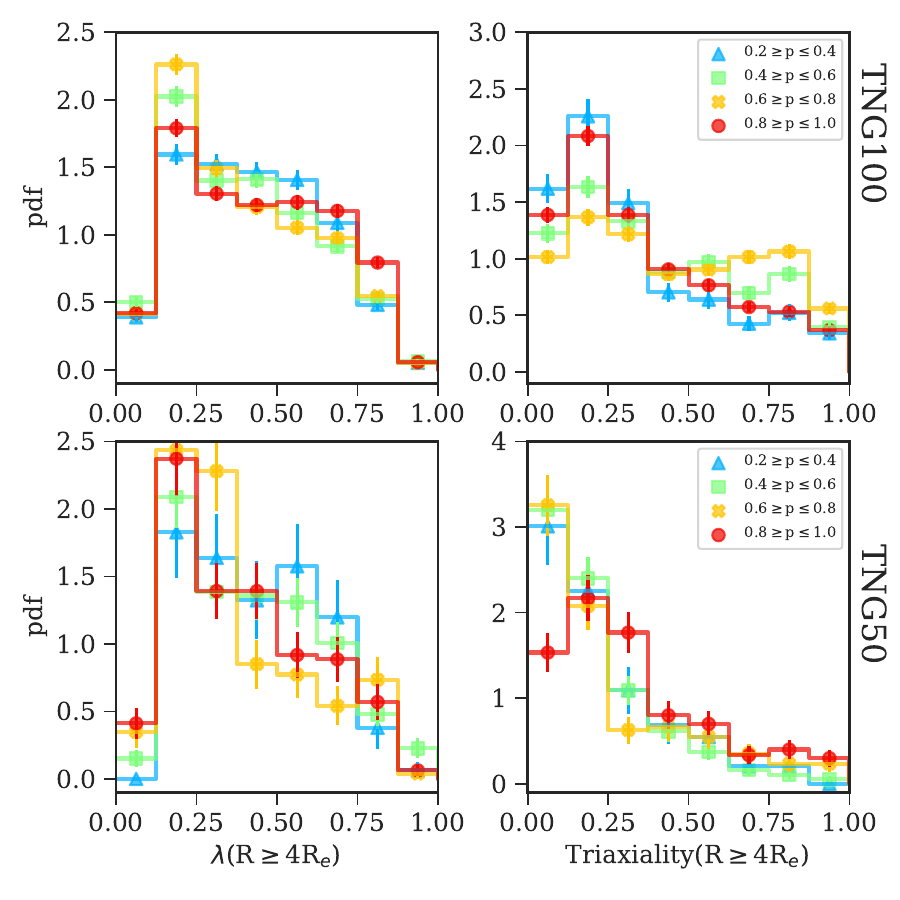}
    \caption{Distribution of $\lambda$ parameter and triaxiality in the stellar halos (i.e. at $R\geq4\re{}$) for different intrinsic shapes in the central regions, parametrized by the intermediate to major axis ratio $p(1\re{})$. TNG100 and TNG50 galaxies are shown separately, in the top and bottom panels respectively. These distributions are not systematically dependent on $p(1\re{})$,  and thus the stellar halo spin and triaxiality remain unbiased after implementing a sample selection based on $p(1\re{})$.}
    \label{fig:bias_in_halo_props}
\end{figure}

The middle row of Fig. \ref{fig:lambda_ell_center} shows that the final selected sample of ETGs lies in the region of the diagram populated by the observed galaxies. The fraction of simulated galaxies in the region of avoidance (i.e. above the black and below the magenta lines) is in agreement with the $\lambda_e-\varepsilon$ observations. The location of the simulated galaxies in the plane follows closely the Atlas3D, SAMI, and MASSIVE galaxies. In the MANGA sample there is a large fraction S0 galaxies with $\lambda_e>0.7$ that are not present in the other surveys, and are likely due to differences in the data analysis, possibly to the beam corrections applied by \citet{2018MNRAS.477.4711G} on the MANGA data \citep[see discussion in ][]{2019A&A...632A..59F}. 

Figure \ref{fig:bias_in_halo_props} demonstrates that the distribution of stellar halo properties which we are interested in, i.e. $\lambda$ and triaxiality parameter, are not affected by the sample selection based on $p(1\re{})$. The distributions do not systematically depend on the intrinsic shape of the central regions of the galaxies. As discussed in a companion paper, the properties of the galaxies at large radii are mainly set by their accretion history and not by the details of the star formation in the central regions of galaxies. 

The bottom panel of Fig. \ref{fig:MassFunction} shows the stellar mass function for the final sample of ETGs, compared with observations. The stellar mass function of the TNG100 ETGs is still similar to Atlas3D. For TNG50 the additional selection has excluded a large fraction of red galaxies in the stellar mass range $\sim 10^{10.3} - 10^{11} \MSUN{}$. This results in a stellar mass function skewed towards high masses (and so more similar to ePN.S).

Henceforth we classify galaxies as slow rotators (SRs) and fast rotators (FRs), using the dividing line introduced by \cite{2011MNRAS.414..888E},
\begin{equation}
  \lambda_e = 0.31 \sqrt{\varepsilon_e},
  \label{eq:fast_slow_def}
\end{equation}
shown in Fig.~\ref{fig:lambda_ell_center} with the black line: galaxies above this threshold are FRs, and galaxies below are SRs. 
To reduce the effects of inclination, we choose to classify the simulated galaxies using the values of $\lambda_e$ and ellipticity for their edge-on projection (shown in Fig.~\ref{fig:lambda_ell_edgeon}).

\subsection{Summary of the sample selection criteria}\label{sec:sample_selection_summary}

The sample of ETG galaxies used in the remainder of this paper is extracted from the TNG50 and TNG100 simulations by 
\begin{itemize}
    \item  selecting galaxies in the stellar mass range $10^{10.3} \leq M_{*} \leq 10^{12} \MSUN{}$ and with red $(g-r)$ color as in Eq. \eqref{eq:color_mass_sel} (see Fig. \ref{fig:SelectionSample});
    \item excluding a small number of objects with $\re<2\rsoft{}$, to assure sufficient resolution at $1\re{}$;
    \item finally, removing a class of centrally elongated, triaxial galaxies with $p(1\re{})<0.6$, which are systems not present in the observed ETG samples, that probably became bar-unstable and quenched during the process of (central) disk formation. 
\end{itemize}

The selected sample has a distribution of $\lambda_e-\varepsilon(1\re{})$ similar to observed ETGs (Fig. \ref{fig:lambda_ell_center}) and halo properties that are unbiased by the selection in intrinsic shape (Fig. \ref{fig:bias_in_halo_props}).
The mass functions of the selected TNG50 and TNG100 ETG samples are given in Fig.~\ref{fig:MassFunction}, and the mass-size relations are shown in Fig.~\ref{fig:MassSize}.


\section{Photometric properties of the TNG ETG samples}\label{sec:Photometry}  

In this section we study the photometric properties of the selected sample of TNG galaxies and how they vary with radius. Section~\ref{sec:measuring_Re} discusses the measured galaxy sizes and how our definition of effective radii compares with effective radii inferred from ETG photometry. Section~\ref{sec:ellipticity_distribution_center} compares the distribution of projected ellipticities at $1\re{}$ with that from ETG surveys and validates our sample selection. Section~\ref{sec:ellipticity_profiles} studies the TNG ellipticity profiles out to the stellar halo, Sect.~\ref{sec:halo_intrinsic_shape_distr} explores the intrinsic shape distribution of stellar halos and its dependence on stellar mass, and Sect.~\ref{sec:triaxiality_profiles} investigates the dependence of galaxy triaxiality on radius and stellar mass in the simulated samples. Finally Sect.~\ref{sec:signatures_triaxiality_photometry} tests the ability of photometric twist measurements to establish the underlying triaxiality in the TNG galaxies.

\subsection{Sizes of the TNG galaxies} \label{sec:measuring_Re}

We first discuss the adopted measurement of the effective radius for the simulated galaxies, which we will use in the paper as galactocentric distance unit. 

\begin{figure}  
    \centering
    \includegraphics[width=\linewidth]{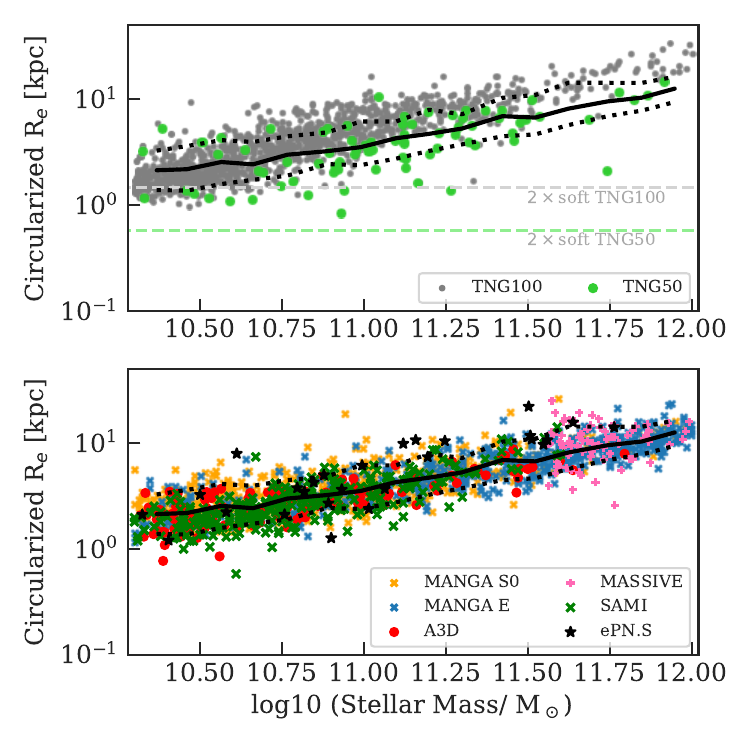}
    \caption{Circularized effective radii, i.e. $\re{}\sqrt{1-\varepsilon}$, of the selected ETG samples in the TNG100 and TNG50 simulations as a function of stellar mass, and comparison with observations (bottom panel). The black curves show the median profile of the distribution of galaxy $\re{}$ from all surveys and the 10th and 90th percentiles, in both panels. 
    The dashed lines in the top panel indicate the resolution of the two simulations.} 
    \label{fig:MassSize}
\end{figure} 

The effective radius \re{} is derived for each projection (edge-on or random LOS) of the galaxies by using cumulative mass profiles in elliptical apertures: \re{} is the major axis radius of the aperture that contains half of the total bound stellar mass. 

Figure \ref{fig:MassSize} shows the circularized \re{} as a function of $M_{*}$ for the final samples of ETGs, and compares it to the distribution of observed effective radii from the different surveys. The \re{} in TNG100 are larger than most of the observed $\re{}$ at $M_{*}\gtrsim10^{10.75}$, but they are in reasonable agreement with the ePN.S measurements.  TNG50 produces smaller galaxies compared to TNG100 and to observations at intermediate stellar masses \citep{2019MNRAS.490.3196P}. This is purely a resolution effect as discussed in Sect.~\ref{sec:simulations}. On the other hand, comparisons to observed \re{} strongly depend on the operational definitions of galaxy sizes, as discussed by \cite{2018MNRAS.474.3976G}.

Observers measure \re{} by integrating light profiles fitted to the bright central regions to large radii. This definition of \re{} tends to underestimate the size (and at the same time the total stellar mass) of the galaxies if the photometric data are not deep enough to sample the light distribution in the halos, especially in massive galaxies with high S{\'e}rsic indices.
\cite{2018A&A...618A..94P} determined \re{} of the ePN.S galaxies from the most extended photometric profiles available in the literature, using a S{\'e}rsic fit of the {\sl outermost regions} to integrate to large radii.  This approach leads to an average increase of the \re{} by a factor of $\sim2$ for the most massive objects with $M_{*}>10^{11}\MSUN{}$. However it does not take into account the possibility of an extra halo component/intra-group or intra-cluster light (ICL) at large radii. For the simulated galaxies, defining the stellar content of the galaxies as all the bound stellar particles identified by the {\sc subfind} algorithm, automatically includes also ICL stars in the most massive halos, thus overestimating both \re{} and $M_{*}$. 

To quantify these effects requires separating a galaxy from the surrounding ICL. A kinematic separation of the ICL similar to \cite{2015A&A...579A.135L} is beyond the scope of this paper. However, \cite{2020ApJS..247...43K} recently found that if the ICL component in bright cluster galaxies is identified as the outer component of a double S{\'e}rsic fit, the radius at which it starts dominating is $\sim 100$ kpc with a very large scatter (5 to 400 kpc in their figure 16). We evaluated the differences in \re{} and $M_{*}$ that we would obtain if instead of using the whole bound stellar mass we limit the galaxy to the mass within 100 kpc. We find that in TNG100 galaxies with $M_{*}<10^{10.5}$ the effects are negligible; in $10^{10.5}\MSUN{}<M_{*}<10^{11}\MSUN{}$ objects the differences in the derived \re{} and stellar masses are within 10\% and 5\% respectively, while between $10^{11}\MSUN{}<M_{*}<10^{11.5}\MSUN{}$ they are within 30\% and 15\%. At higher masses the differences in \re{} can be larger than 50\% and those in $M_{*}$ larger than 25\%, with a very large scatter. These effects are half as pronounced in TNG50. A size-stellar mass diagram analogous to Fig. \ref{fig:MassSize} using the 100 kpc aperture instead of the total bound mass shows an improved agreement with the observed $\re{}$, but TNG100 galaxies with $M_{*}>10^{10.75}$ are still larger on average. This may indicate that TNG100 predicts too large sizes for high mass galaxies \citep[see also][]{2018MNRAS.474.3976G}.

Because of the somewhat arbitrary choice of the 100 kpc limit on one hand, and the uncertainties in the observed $\re{}$ distribution on the other (from differences in sample selection, quality of the photometric data, definition of total stellar light, the mass-to-light ratio to obtain total stellar masses), we choose here to define \re{} for the simulated galaxies as the half mass radius of the total bound stellar mass, and consider the above uncertainties in the discussion of the results where relevant.

\begin{figure*}[ht] 
    \centering
    \begin{minipage}{.5\textwidth}
        \centering
        \includegraphics[width=\linewidth]{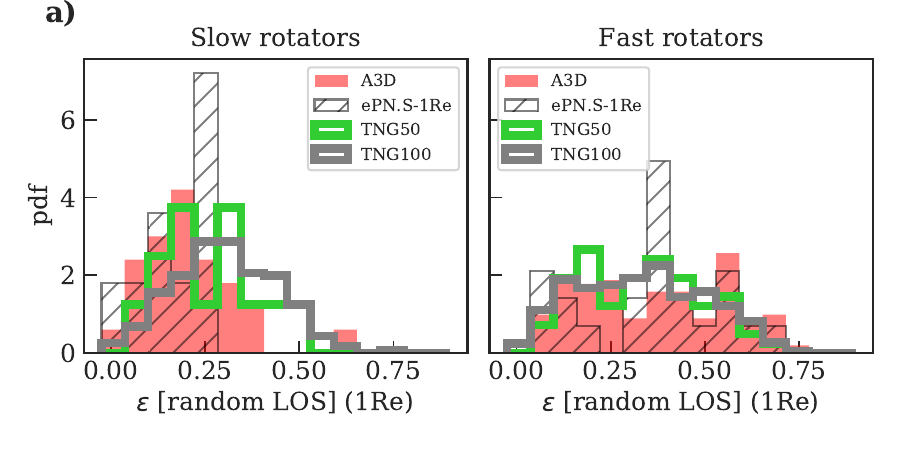}
        \includegraphics[width=\linewidth]{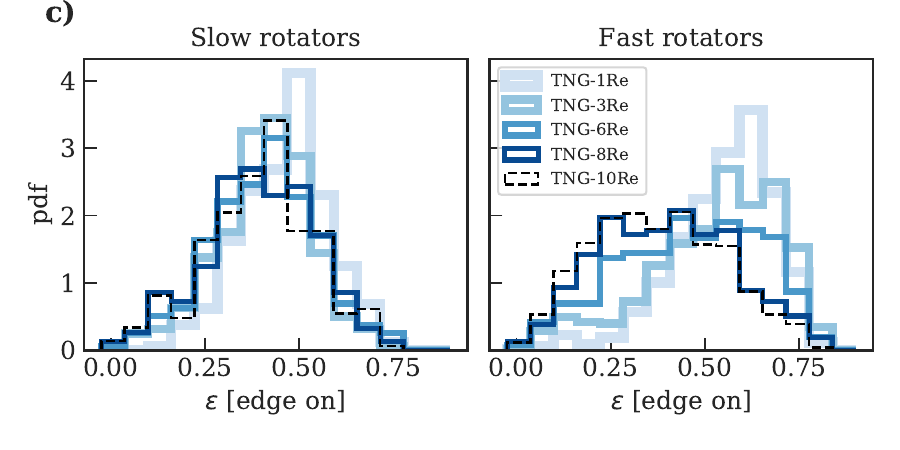}
    \end{minipage}%
    \begin{minipage}{0.5\textwidth}
        \centering
        \includegraphics[width=\linewidth]{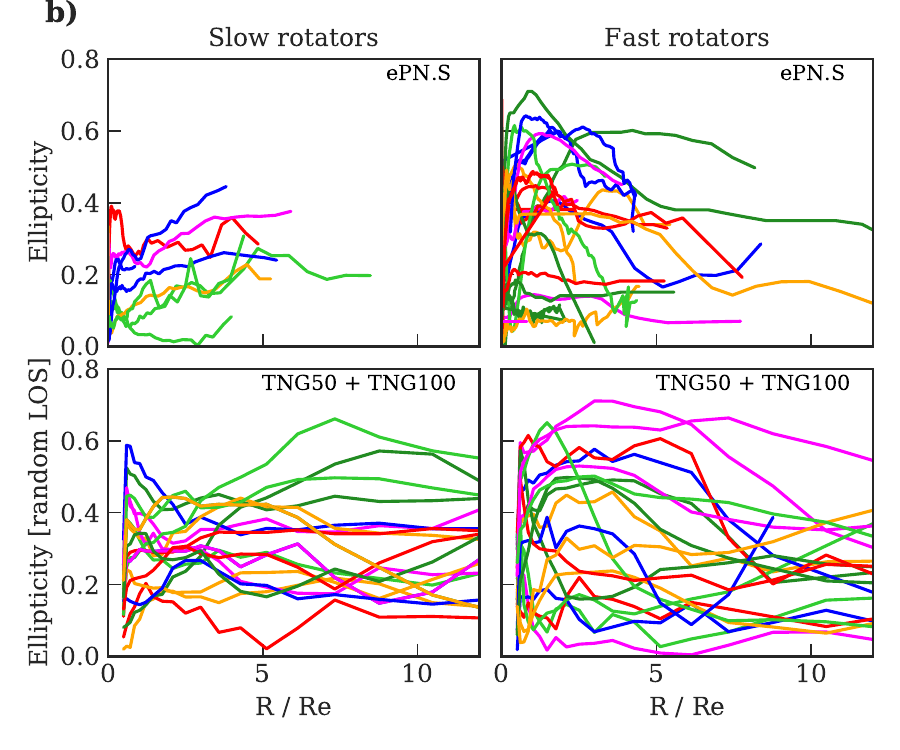}
    \end{minipage}
    \caption{Ellipticity distribution and profiles of the selected TNG ETGs. \textbf{a)} Distribution of the random LOS $\varepsilon(1\re{})$ compared with Atlas3D and ePN.S as indicated in the legend. Despite some differences between the Atlas3D and the TNG SRs, the good agreement of FR distributions shows that the selected sample of TNG galaxies contains a mixture of disk and spheroidal galaxies consistent with Atlas3D.
    \textbf{b)} Ellipticity profiles for the ePN.S galaxies (top) and 10 randomly selected example galaxies from TNG100 and 10 from TNG50, projected along a random LOS (bottom). The comparison between TNG and ePN.S galaxies highlights the variety of projected ellipticity profiles in both samples.
    \textbf{c)} Distribution of the edge-on projected ellipticity values at different radii predicted by TNG. SRs have a rather constant ellipticity distribution with radius. The FRs have a large variety of shapes at large radii as shown by the broadening of the distribution, while the shift of the peak to lower ellipticities shows the tendency of most galaxies to become rounder in their halos. }
    \label{fig:ElliptDistr}
\end{figure*}

\subsection{Ellipticity distribution in the central regions}
\label{sec:ellipticity_distribution_center}

Figure \ref{fig:ElliptDistr}a shows the distributions of the ellipticities measured at 1 \re{} for the final sample of selected ETGs, compared with Atlas3D and ePN.S. In the top panels are the SRs, and in the bottom the FRs. 

The TNG50 and TNG100 simulations predict a significant fraction of SR galaxies with $\varepsilon(1\re{})>0.4$, while the observed SRs are relatively rounder. This is a common feature of current simulations \citep{2014MNRAS.444.3357N, 2018MNRAS.480.4636S}, and its origin is still to be understood. In the case of the FR class, the ellipticity distributions are rather flat-topped, and in good agreement with Atlas3D. 
By comparison the ePN.S sample contains on average rounder (and also more massive, see the bottom panel of Fig. \ref{fig:MassFunction}) galaxies, and hence a lower number of disk galaxies: none of the ePN.S FRs has ellipticity higher than 0.7.  

Overall Fig. \ref{fig:ElliptDistr}a shows that the selected sample of ETGs contains a mixture of galaxy types consistent with observations, with a similar balance between disks and spheroids.

\subsection{Ellipticity profiles}
\label{sec:ellipticity_profiles}

The ellipticity profiles of the TNG galaxies are compared over an extended radial range with those of the ePN.S galaxies in Fig.  \ref{fig:ElliptDistr}b. There we show profiles for randomly selected sub-samples of the simulated galaxies. Figure \ref{fig:ElliptDistr}c instead shows the distribution of ellipticities at different radii for the fast and the slow rotators separately.

The observed profiles for the ePN.S SRs generally mildly increase with radius, reaching $\varepsilon\sim0.3$ at $4\re{}$. By comparison, the simulated SRs have more nearly constant ellipticity profiles. This can also be seen in the histograms of Fig. \ref{fig:ElliptDistr}c where the $\varepsilon$ distribution is almost unvaried between different radii. 

Most of the simulated FRs have decreasing ellipticity profiles with radius, while a fraction have high ellipticity also at large radii, as also shown by the ePN.S galaxies. Thus, Fig. \ref{fig:ElliptDistr}c shows that at larger radii the FR ellipticity distribution peaks at smaller $\varepsilon$ and, at the same time, it broadens.

The decrease in ellipticity of the majority of the FRs supports the idea of a change in structure of these galaxies at large radii. The large range of flattening in the stellar halos indicates a variety in the stellar halo properties. By comparison, the SRs show only small structural variations.

\begin{figure}  
    \centering

    \includegraphics[width=0.9\linewidth]{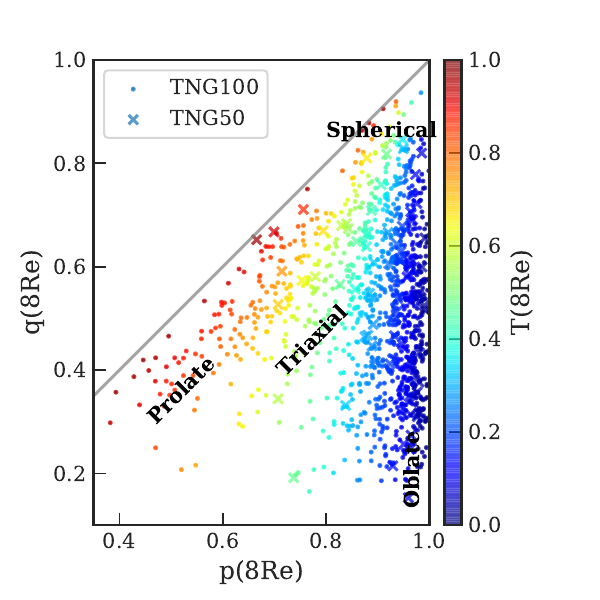}    \includegraphics[width=0.9\linewidth]{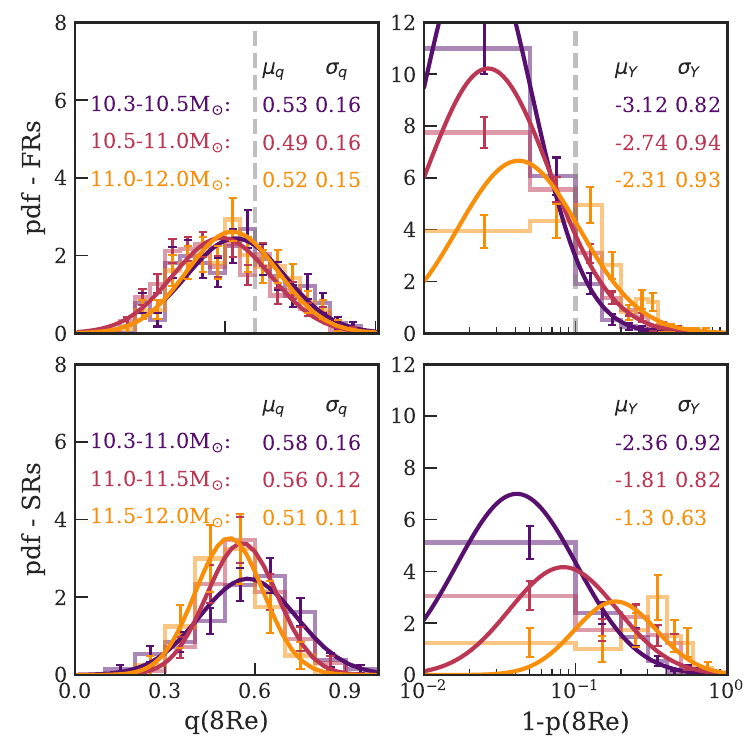}
    \caption{Intrinsic shape distribution of ETG stellar halos in TNG. \textbf{Top}: minor to major axis ratio $q$ versus intermediate to major axis ratio $p$ coloured by triaxiality as measured at 8\re{}. TNG50 and TNG100 galaxies are shown with different symbols as in the legend.
    \textbf{Bottom}: Intrinsic shape distribution for the halos ($r\sim8\re{}$) of the FRs (top panels) and SRs (bottom panels) in mass intervals, shown with different colors as indicated in the figure. The fitted functions are shown with solid lines, and the fitted parameters are reported in the legend. The vertical dashed lines show the comparison with the photometric model used in \cite{2018A&A...618A..94P} to reproduce the observed photometric twists and average ellipticities for the ePN.S survey. Most low-mass TNG galaxies have near-oblate stellar halos (top), changing towards more triaxial shapes with increasing stellar mass (bottom panels).}
    \label{fig:intrinsic_shapes_distributions_stars}
\end{figure}

\subsection{Intrinsic shape distribution of the stellar halos}
\label{sec:halo_intrinsic_shape_distr}

\begin{figure*}[ht]  
    \centering

    \includegraphics[width=\linewidth]{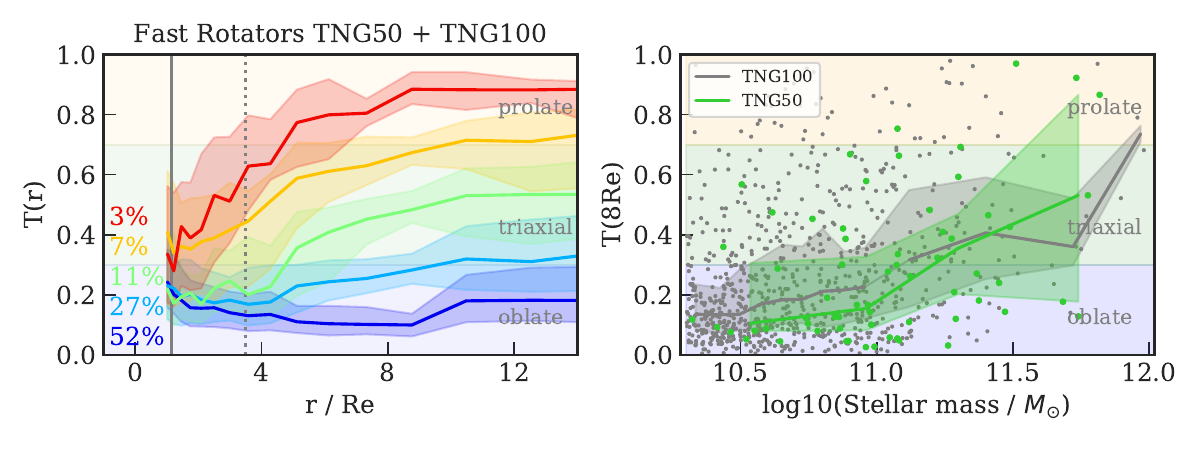}
    \includegraphics[width=\linewidth]{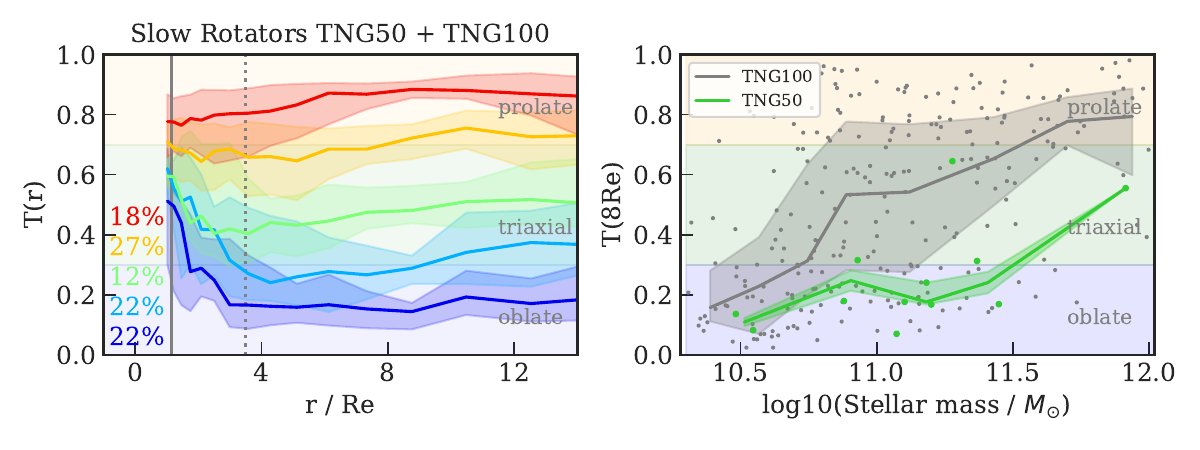}

    \caption{\textbf{Left panels}: Median triaxiality profiles for the fast rotators (top) and the slow rotators (bottom). The numbers on the left indicate the percentage of FRs or SRs populating each profile. The vertical dotted lines show $r=9\rsoft{}$ for TNG100 galaxies with $M_{*}\sim10^{10.3}M_{\odot}$, the solid lines show $r=9\rsoft{}$ for TNG100 galaxies with $M_{*}\sim10^{11}M_{\odot}$. At radii larger than these the profiles are not affected by resolution issues.
    \textbf{Right panels}: The median of the stellar halo triaxiality distribution measured at 8\re{} as a function of stellar mass. The shaded areas are contoured by the 25th and 75th percentiles of the distribution. TNG100 and TNG50 are shown separately, FRs are on top, SRs on the bottom. 
    On average the triaxiality parameter increases with radius and with stellar mass. The FRs have increasing $T(r)$ profiles, while SRs have either constant or decreasing profiles.}
    \label{fig:triaxiality_profiles}
\end{figure*}

In this section we quantify the distribution of the galaxy intrinsic shapes at large radii. Here we refer to stellar halo as the outer regions of the galaxies, where the physical properties are markedly different from those of the central regions. While this region may begin at different radii in each ETG, we will see in the next section that the median triaxiality profiles for our sample reach constant values beyond $\sim5\re{}$. Hence we measure the stellar halo intrinsic shape distributions by deriving the intrinsic axes ratios in a shell around 8 \re{}, 1.5\re{} thick, which is the maximal radius at which also the lowest mass systems contain enough particles to reliably measure intrinsic shapes, see Sect.\ref{sec:measuring_IntrinsicShape}. Different choices of the shell thickness, or slightly different choices of the radius (for example 7 instead of $8\re{}$) at which we measure $p$ and $q$ deliver similar results. 

The top panel of Fig.  \ref{fig:intrinsic_shapes_distributions_stars} shows the minor to major axis ratio $q$ as a function of the intermediate to major axis ratio $p$: we find a large variety of possible shapes, from very flat near-oblate with $q\sim0.3$, to prolate with $q\sim p \sim 0.5$. The majority of (low-mass) galaxies appear to have near-oblate stellar halos, with a large scatter in minor to major axis ratio $q$. The bottom panels of Fig.  \ref{fig:intrinsic_shapes_distributions_stars} show the intrinsic shape distributions for the fast and slow rotators separately. The distributions of the minor to major axis ratio $q$ resemble Gaussians,  and fitted as such the FRs have mean $\mu_q \sim 0.5$ and dispersion $\sigma_q\sim0.16$ in all stellar mass bins. The SRs have $\mu_q\sim0.6$ and $\sigma_q\sim0.15$, with a tendency for the highest mass galaxies to be flatter. 

The distribution of the intermediate to major axis ratio $p$ can be approximated by a log-normal distribution in $Y = \ln(1-p)$. The shape of this distribution shows a dependence on stellar mass: at higher stellar masses $\mu_Y$ increases, together with the width of the distribution. This means that at higher stellar masses, in both the FR and SR classes, the fraction of near-oblate galaxies with $p\sim1$ decreases.

The vertical dashed lines in Fig.  \ref{fig:intrinsic_shapes_distributions_stars} shows the comparison with the photometric model used by \cite{2018A&A...618A..94P} to reproduce the distribution of maximum photometric twists versus mean ellipticity of the observed FRs. Their model parameters $\mu_q=0.6$ and $\mu_p=0.9$ are within $1\sigma$ of the mean values obtained from the distribution of simulated FRs.

\subsection{Triaxiality profiles}
\label{sec:triaxiality_profiles}

We can study how the intrinsic shapes of galaxies vary as a function of radius by looking at their triaxiality profiles. We recall from Sect. \ref{sec:measuring_IntrinsicShape} that because of the error due to resolution effects in the central regions, $T$ profiles are considered well-defined only beyond $r=9\rsoft$, where their error $\Delta T\sim0.2$ for typical FR axis ratios (App.\ref{appendix:accuracy_shapes_measurement}).  Thus we show profiles only for $r> 3.5\re{}$ for the lowest mass objects, and for $r>1.16\re{}$ for galaxies with $M{*}\sim10^{11}\MSUN$.

The left panels of Fig.  \ref{fig:triaxiality_profiles} show the median triaxiality profiles for FRs and SRs.  These median profiles were built by binning the galaxies according to their values of the triaxiality parameter T at $8\re{}$, and are plotted against the intrinsic major axis distance $r$. The scale radius $\re{}$ that normalizes the three dimensional radius is the 2D projected effective radius for the edge-on projection of each galaxy. 
The right panels show the median of the distribution of the triaxiality parameter measured at 8\re{} as a function of the stellar mass. 

FRs are characterized by increasing $T$ profiles, which tend to plateau at $r > 5\re{}$ where the TNG galaxies show a broad range of intrinsic shapes despite all having near-oblate centers. SRs tend to have flatter profiles. 

We find that the stellar halo intrinsic shape distribution is a function of stellar mass. This is visible in the right hand side of Fig.  \ref{fig:triaxiality_profiles}, for FRs and SRs separately. 
At lower masses the TNG galaxies have preferentially near-oblate shapes, with $T\lesssim 0.2$, but at larger masses the median triaxiality parameter increases, so that at $M_{*}>10^{11}\MSUN{}$ there is a non-negligible fraction of galaxies with prolate-triaxial halos, even among the FRs. 
We note a systematic difference between TNG50 and TNG100 in the triaxiality of the SR stellar halos. In TNG50 the SRs tend to be much more oblate, indicating some higher degree of dissipation involved in their evolution compared to the SRs in TNG100. The statistical significance of this difference is marginal since TNG50 contains only 14 SRs.

\subsection{Photometric twists and triaxiality in TNG ETGs} \label{sec:signatures_triaxiality_photometry}

\begin{figure}  
    \centering
    \includegraphics[width=\linewidth]{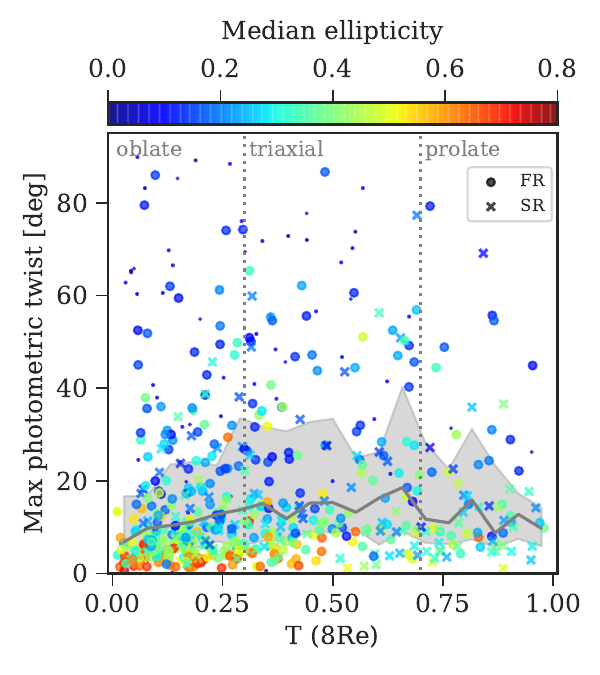}  
    \caption{Maximum measured photometric twist between 1 and 8 \re{} as a function of the halo triaxiality $T(8\re{})$ for the TNG50 and TNG100 galaxies projected along a random LOS. FRs and SRs are shown with different symbols as in the legend. Smaller symbols are used to represent galaxies with ellipticity $<0.1$, for which the errors on $\PAphot{}$ are large. The color-bar indicates the median ellipticity measured between 1 and 8\re{}. The gray solid line shows the median of the photometric twist distribution as a function of $T(8\re{})$; the shaded area encloses the 25th and 75th percentiles. The amplitude of the photometric twists in TNG galaxies is only weakly dependent on the triaxiality parameter.}
    \label{fig:Phot_twist}
\end{figure}

Isophotal twists in photometry are generally considered to be signatures of triaxiality. This is because the projection on the sky of coaxial triaxial ellipsoids with varying axis ratios approximating the constant luminosity/mass surfaces of ETGs can result in twisting isophotes \citep[e.g.][]{1980MNRAS.193..885B}. However,
the effects of triaxiality on the $\PAphot{}$ profiles are model dependent, that is they depend on axis ratio, on how much the axis ratios changes with radius, as well as on the viewing angles.
 
Figure \ref{fig:Phot_twist} shows the distribution of maximum photometric twist, i.e the maximum variation of $\PAphot{}(R)$, measured between 1 and 8 \re{}, as a function of the halo triaxiality at 8 \re{}, i.e. where the triaxiality profiles have reached a constant value. Each symbol in the diagram is color coded by the median projected ellipticity between 1 and 8 \re{}.  
Galaxies with ellipticity lower than 0.1 have the photometric position angle poorly determined, and are shown with smaller symbols. 

We observe that the amplitude of the photometric twists is only weakly dependent on the triaxiality. Near-oblate and near-prolate galaxies are slightly less likely to have constant $\PAphot{}$ than triaxial galaxies, but the majority of the galaxies have small twists irrespective of $T(8\re{})$. This is explained by the fact that large twists can be measured for viewing angles close enough to face-on \citep[][that is lower ellipticities in Fig. \ref{fig:Phot_twist}]{2018A&A...618A..94P}, at which even small values of $T$ can produce large twists. 
From Fig. \ref{fig:Phot_twist} we conclude that the amplitude of the photometric twists is a poor indicator for galaxy triaxiality, and that very small photometric twists are intrinsically compatible with triaxial shapes.


\section{The kinematics properties}\label{sec:Kinematics}

In this section we study how the kinematic properties of the TNG galaxies vary with radius. In Sect. \ref{sec:lambda_profiles} we derive median differential $\lambda(R)$ profiles to quantify the variety of kinematic behaviors in the outskirts of FRs and SRs. Sections \ref{sec:simulated_VS_observed_rotation_center} and \ref{sec:simulated_VS_observed_rotation_halo} compare the shapes of the $V_\mathrm{rot}/\sigma$ profiles of the simulated ETGs with the observed galaxies in the Atlas3D and ePN.S surveys. 
Finally Sect. \ref{sec:signatures_triaxial_halos_in_kinematics} uses the simulated galaxies to assess kinematic misalignments and twists as signatures of triaxial shapes in galaxies.

\subsection{Lambda profiles}\label{sec:lambda_profiles}

\begin{figure*}[ht]  
    \centering

    \includegraphics[width=1\linewidth]{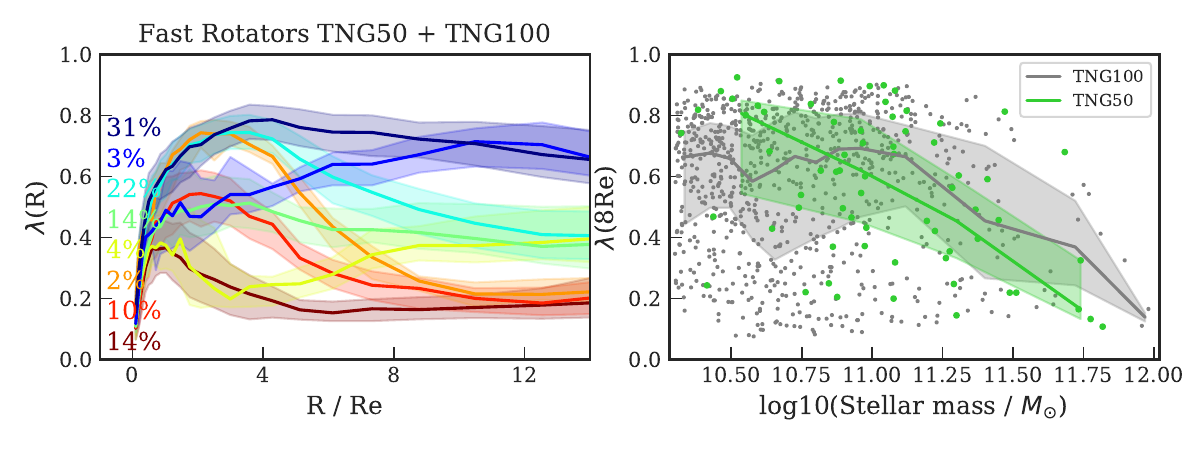}
    \includegraphics[width=1\linewidth]{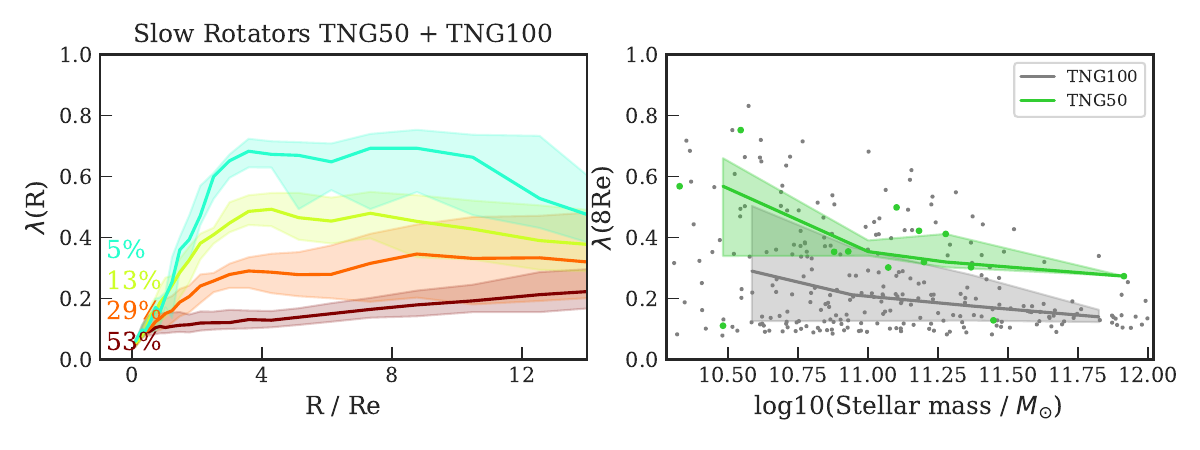}

    \caption{\textbf{Left}: Median $\lambda$ profiles for fast rotators (upper left) and slow rotators (lower left panel) in their edge on projection. The median profiles are built by binning the galaxies according to the shape of their $\lambda$ profiles; see text. The shaded areas are bounded by the 25th and 75th percentiles of the distributions. The numbers on the left indicate the percentage of FRs or SRs populating each profile. \textbf{Right}: Distribution of $\lambda(8\re{})$ for the full sample, as a function of stellar mass. FRs and SRs are shown separately, as well as TNG50 and TNG100. FR galaxies show a large variety of $\lambda$ profiles, whereas SRs have either constant or increased $\lambda$ in the stellar halo. Overall the halo rotational support decreases at high stellar masses. }
    \label{fig:lambda_profiles}
\end{figure*}

The top panels of Fig.  \ref{fig:lambda_profiles} show the median differential $\lambda$ profiles for FRs and SRs in their edge-on projection. 
Galaxies are binned together according to the shape of their profiles. We achieve this by binning the FRs according to their values of $\lambda$ at $R=4\re{}$ and at $R=8\re{}$. For the SRs the shape of the $\lambda(R)$ profiles is generally a monotonic function of $R$: in this case we binned the profiles according to their $\lambda(7\re{})$. 

Most of the FRs reach their maximum $\lambda(R)$ around 3\re{}; only 7\% of the galaxies increase $\lambda(R)$ between 3 and 10 \re{}. FRs divide almost evenly among a third (34\%) that have flat $\lambda$ profiles with $\lambda(8\re{})>0.6$, a third (40\%) with gently decreasing profiles and $0.3>\lambda(10\re{})\geq0.6$, and another third (26\%) with very low rotation in the halo ($\lambda(10\re{})\leq0.3$).

The SRs essentially divide between a half (53\%) with non-rotating halos ($\lambda(7\re{})<0.2$) and a half with increased rotational support at large radii compared to the central regions. We observe that a small fraction of the SRs (5\% of the TNG100 SRs and 15\% of the TNG50 SRs) reach very high values of $\lambda$ at large radii ($\lambda(7\re{})>0.6$).
The majority of these galaxies are genuine slow rotators with strongly rotating halos similar in terms of velocity fields and $\lambda$ profiles to observed SRs like NGC 3608 \citep{2018A&A...618A..94P}. The others are galaxies with a clear extended disk structure characterized by rapid rotation and low velocity dispersion, but whose central kinematics is dominated by a non-rotating bulge. There are no observed counterparts for the latter in both the ePN.S (33 galaxies) and the SLUGGS surveys \citep[][25 galaxies, of which 18 are in common with ePN.S]{2016MNRAS.457..147F}. A larger sample of galaxies with extended kinematic data would be needed to confirm or rule out these objects.

For the SRs we note a mismatch in the amount of halo rotational support between TNG50 and TNG100, analogous to the difference observed for the halo triaxiality (Sect. \ref{sec:triaxiality_profiles}): on average TNG50 SR halos rotate faster and have more oblate shapes. These differences in halo properties might be due to the dependence of the galaxy formation model on the resolution of the simulations, although the number of SRs in TNG50 (only 14 galaxies) is too small to draw strong conclusions.

For both FRs and SRs, and in both TNG50 and TNG100, the stellar halo rotational support depends weakly on stellar mass up to $M_{*}\sim10^{11.3}\MSUN{}$ (Fig. \ref{fig:lambda_profiles}). However, at high stellar masses the fraction of galaxies with significant rotation in the halo decreases, so that at $M_{*}>10^{11.5}\MSUN{}$ most of the galaxies have non rotating halos. The broad range of possible $\lambda(R)$ profile shapes in Fig.  \ref{fig:lambda_profiles} shows that the IllustrisTNG simulations encompass, if not exceed, the observed variety of halo rotational support found in the ePN.S survey, of which one of the key results was the large kinematic diversity of stellar halos.

\subsection{Simulated versus observed rotation profiles - central regions} \label{sec:simulated_VS_observed_rotation_center}

\begin{figure}  
    \centering
    \includegraphics[width=\linewidth]{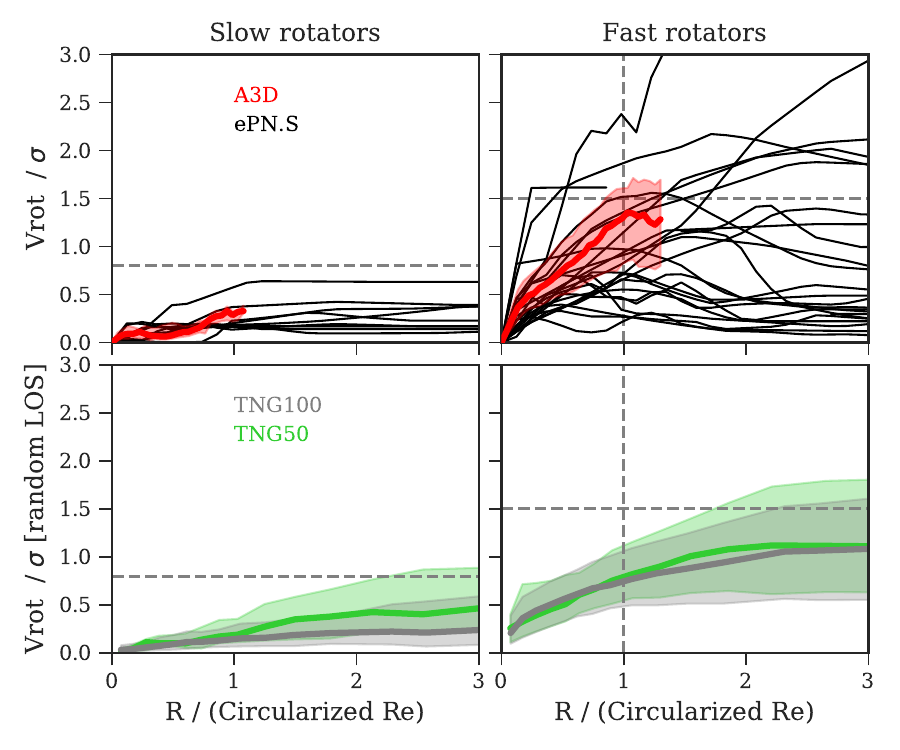}
    \caption{Median $V_\mathrm{rot} / \sigma$ profiles for slow rotators (left) and fast rotators (right panels). \textbf{Top}: observations from Atlas3D and ePN.S. For the former we show median profiles and shaded  areas bounded by the 20th and the 80th percentiles of the distribution. For the ePN.S galaxies we show individual profiles.  \textbf{Bottom}: median profiles and 20\%-80\% distributions for the TNG50 and TNG100 ETGs, projected along a random LOS.  Vertical and horizontal dashed lines are for guiding the eyes in comparing top with bottom panels.
    This shows that the average $V_\mathrm{rot} / \sigma$ profiles of the simulated ETGs are shallower than the observed profiles. }
    \label{fig:comparisonA3D}
\end{figure}

Figure \ref{fig:comparisonA3D} shows the median $V_\mathrm{rot} / \sigma$ profiles of the TNG100 and TNG50 galaxies compared with median profiles from Atlas3D and individual galaxy profiles from the ePN.S sample for $0\leq R \leq 4 \re{}$. Here we normalize the radii by the circularized $\re{}$, i.e. $\re \sqrt{1-\epsilon(1\re)}$, for an appropriate comparison with the Atlas3D $\re{}$.

The profiles of the simulated SRs are similar to the observed profiles.
The FRs instead show a difference in how quickly $V_\mathrm{rot} / \sigma$ rises with radius: observed FR galaxies have on average more steeply rising $V_\mathrm{rot} / \sigma$ profiles than the simulated ETGs. Very few TNG FRs reach $V_\mathrm{rot}/\sigma \geq 1$ within $1\re{}$ compared to the Atlas3D galaxies, and almost none exceeds $V_\mathrm{rot}\sigma (1\re) = 1.5$ in either TNG100 or TNG50.

The different shapes of the $V_\mathrm{rot}/\sigma$ profiles in observations and simulations cannot be explained with resolution effects, as in {\sl both} TNG50 and TNG100 the $V_\mathrm{rot} / \sigma$ profiles tend to peak at a median radius of $R\sim3\re{}$. By comparison, the ePN.S FRs tend to peak at smaller fractions of \re{}, at a median 1.3\re{}. 

This difference between the observed and simulated FRs is not a consequence of the selection functions of the samples. The TNG galaxies are selected according to color, mass and intermediate to major axis ratio $p$. The selection in $p$ removes centrally elongated galaxies, most of which have intermediate to low $\lambda(1\re)$ (Fig. \ref{fig:lambda_ell_center}). The Atlas3D sample, selected as described in Sect. \ref{sec:observed_quantities}, has some of the disk galaxies removed which as in MANGA (Fig.~\ref{fig:lambda_ell_center}) will be mostly located at high $\lambda(1\re)$. We recall that in the comparison we consistently matched the color selection and mass range of the Atlas3D sample to the selection criteria adopted for the TNG galaxies. Thus the TNG sample should in principle contain a larger number of late type galaxies with strong disks, i.e. high $V_\mathrm{rot}/\sigma$, by comparison with Atlas3D. Figure \ref{fig:comparisonA3D} shows instead that the TNG100 and TNG50 ETG samples lack galaxies with high rotational support at $1\re$.

In Sect. \ref{sec:measuring_Re} we discussed that the effective radii used to normalize the radial scales in TNG would be expected to be systematically slightly overestimated compared to the observed \re{} since they are defined using the total bound stellar mass which is often inaccessible in observations. If taken into account, this effect would {\sl increase} the gap between simulated and observed samples. 
On the other hand, since the mass-size relation is roughly reproduced in the simulations (Fig. \ref{fig:MassSize}), the different steepness of the $V_\mathrm{rot} / \sigma$ profiles in observations and simulations implies a different distribution of the angular momentum as a function of radius in the simulated galaxies. This could be due to a too efficient condensation of the gas into stars that does not allow the gas to dissipate and collapse to small enough radii.

\subsection{Simulated versus observed profiles - outskirts}
\label{sec:simulated_VS_observed_rotation_halo}

\begin{figure*}  

    \includegraphics[width=1.\linewidth,right]{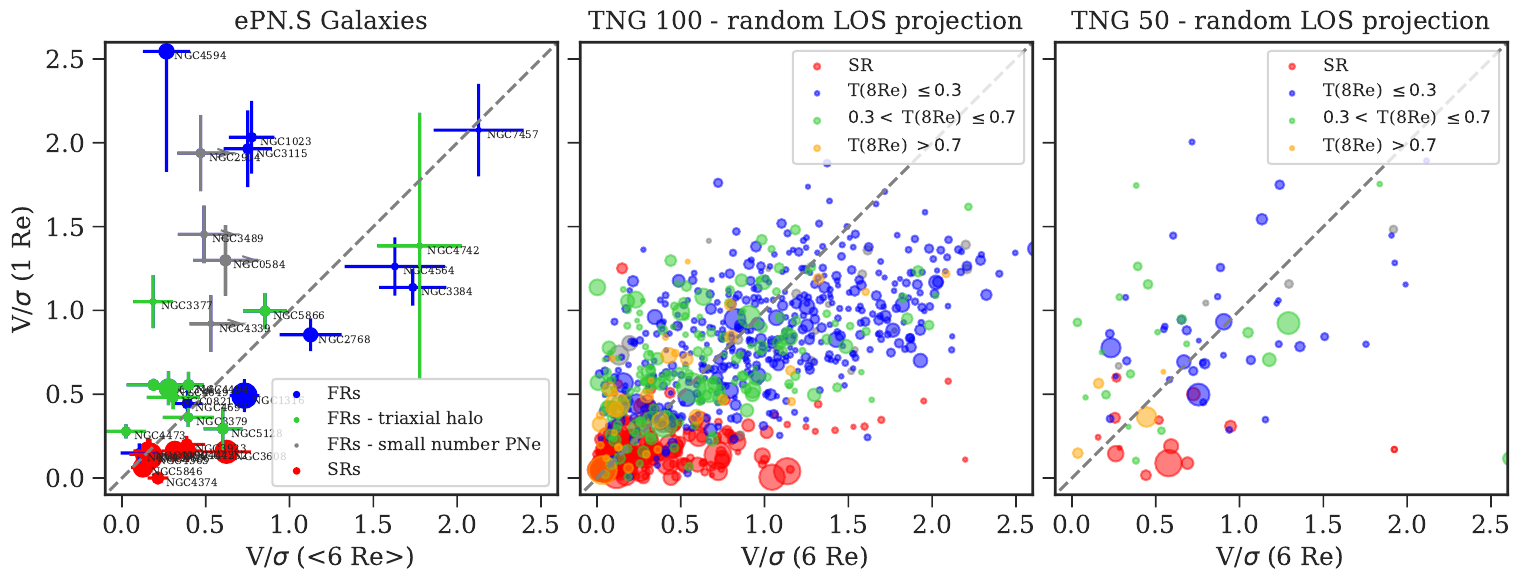}
    \includegraphics[width=0.69\linewidth,right]{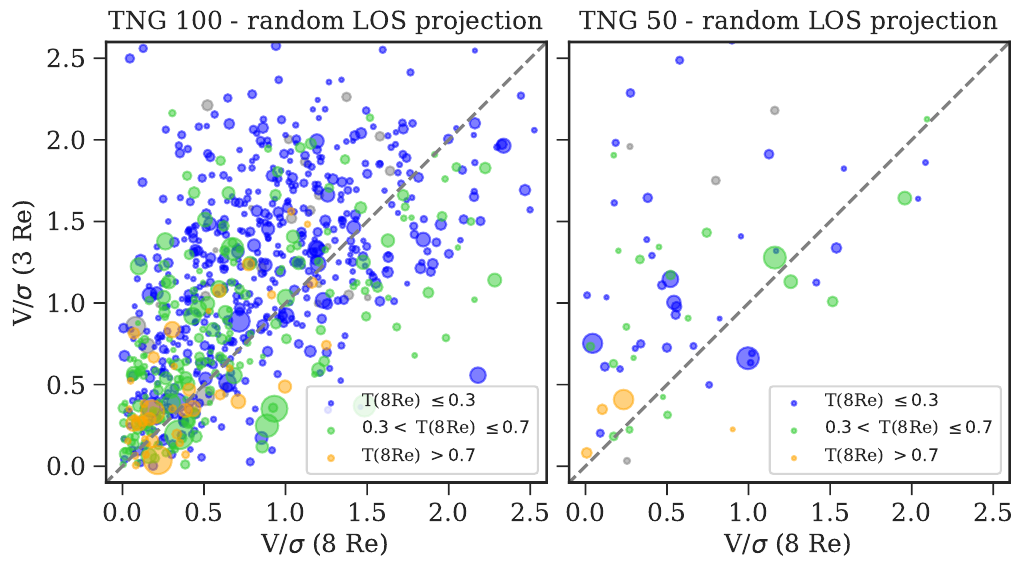}

     \caption{$V/\sigma$ ratio in the central regions compared with the $V/\sigma$ at large radii. \textbf{Top}: $V/\sigma(1\re{})$ versus $V/\sigma(6\re{})$ for the ePN.S galaxies (left), TNG100 (middle), and TNG50 (right). \textbf{Bottom}: $V/\sigma(3\re{})$ versus $V/\sigma(8\re{})$ for TNG100 (left) and TNG50 (right). Different colors in the ePN.S sample mark the SRs (in red), the FRs with kinematic signatures for triaxial halos (i.e. with kinematic twists or misaligments, in green), and FRs with $\PAkin{}$ aligned with $\PAphot{}$ (in blue). The gray points stand for ePN.S FRs for which the measured $V_\mathrm{rot}/\sigma(<6\re{}>)$ is a lower limit. For the simulated galaxies the different colors mark SRs, and FRs with different halo $T(8\re{})$: FRs that have near-oblate halos (in blue); FRs with triaxial halos (in green); near-prolate FRs (in orange). The gray symbols show simulated galaxies with too few particles at $8R_e$ for measuring $T$. The size of the data points is proportional to the stellar mass. The gray dashed lines show the 1:1 relation. The TNG simulations reproduce the diversity of observed ETG halo kinematics and they echo the observed kinematic transitions between the central regions and outskirts of the FR galaxies, albeit at larger radii.} 
    \label{fig:Vsigma}
\end{figure*}

Figure \ref{fig:Vsigma} compares the relation between $V_\mathrm{rot} / \sigma$ in the central regions and $V_\mathrm{rot} / \sigma$ in the stellar halos for the simulated and observed galaxies. The observed galaxies are from the ePN.S survey \citep[][their figure 9]{2018A&A...618A..94P} and are reported in the top left panel. Their measurements use absorption line data at $1 \re{}$ and PN data for the halos, which on average cover $6 \re{}$ with a large scatter (minimum $3\re{}$, maximum $13\re{}$). The central and right top panels show $V_\mathrm{rot}/\sigma(6\re{})$ versus $V_\mathrm{rot}/\sigma(1\re{})$ for TNG100 and TNG50 ETGs, respectively. Galaxies close to the dashed 1:1 lines show similar rotational support in the central regions and in the outskirts; galaxies below the lines have increased rotational support in their stellar halos; galaxies above the lines instead have reduced rotation at large radii.

The position of the observed SRs below the one-to-one line is reproduced by the simulations. As already discussed in Sect. \ref{sec:lambda_profiles} there are a few outliers among the simulated SRs with $V_\mathrm{rot}/\sigma(6\re{}) \gtrsim 1$, some of which are actually extended disks with prominent bulges at the center.  These represent a small fraction of the SR family, and do not have observed counterparts in the ePN.S and SLUGGS surveys.

For the TNG FRs we find a different distribution when comparing rotational support at the same radii:  most of these galaxies fill the bottom half of the diagram, with very few reaching $V_\mathrm{rot}/\sigma(1\re{})>1.5$, and a large fraction having significant $V_\mathrm{rot}/\sigma$ at $6\re{}$. 
This difference is explained by the shallower $V_\mathrm{rot}/\sigma(R)$ profiles of the simulated galaxies compared to the observed FRs, as discussed in Sect. \ref{sec:simulated_VS_observed_rotation_center}. Since the simulated galaxies tend to peak at larger radii than the observed ETGs, there are almost no objects that reach $V_\mathrm{rot}/\sigma > 1.5$ already at $1\re{}$. 

In the bottom panels of Fig. \ref{fig:Vsigma} we show the comparison at adjusted radii: $V_\mathrm{rot} / \sigma(3\re{})$ (i.e. at the median radius where the TNG $V_\mathrm{rot}/\sigma$ profiles tend to reach the maximum) versus  $V_\mathrm{rot}/\sigma(8\re{})$ (i.e. where the decreasing rotation profiles finally reach their minimum, see also Fig. \ref{fig:lambda_profiles}), and find better agreement.
Now the FRs spread in the region of the diagram above the one-to-one line as they do for the ePN.S observations in the top left panel of Fig. \ref{fig:Vsigma}, with a sub-population of objects showing $V/_\mathrm{rot}\sigma(8\re{})>V/\sigma(3\re{})$. We note that FRs with near-prolate stellar halos occupy mostly the lower left corner with low $V_\mathrm{rot}/\sigma(3\re{})$ and low $V_\mathrm{rot}/\sigma(8\re{})$. Galaxies with triaxial halos tend to distribute on the left side of the diagram, galaxies with oblate halos tend to have higher $V_\mathrm{rot}/\sigma(8\re{})$. It is interesting that, aside for the differences in the central regions at $R\lesssim 3\re{}$, the observed galaxies with and without signatures for triaxial stellar halos distribute similarly as the simulated triaxial and near-oblate stellar halos, respectively, with the triaxial halos having on average lower $V_\mathrm{rot}/\sigma(8\re{})$.

From Fig. \ref{fig:Vsigma} we conclude that even though the quantitative details between simulated and observed ETGs galaxies do not agree, the simulations do reproduce the observed kinematic transitions between central regions and outskirts, as well as the variety of halo kinematic classes. 
 
\subsection{Relation of kinematic misalignments and twists with triaxiality in TNG ETGs}
\label{sec:signatures_triaxial_halos_in_kinematics}
\begin{figure}  
    \centering
    \includegraphics[width=\linewidth]{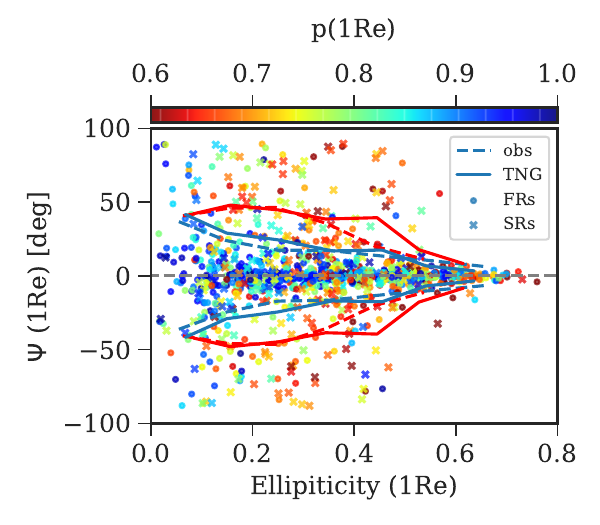} 
    \caption{Distribution of the misalignments $\Psi = \PAkin{}-\PAphot{}$ at $1\re{}$ as a function of the ellipticity for both TNG100 and TNG50 galaxies. Data points are color coded according to their intermediate to major axis ratio $p(1\re)$; open symbols are SRs, filled symbols are FRs. The solid lines show the scatter of the misalignments as a function of the ellipticity for the TNG galaxies, calculated on the mirrored $(\Psi,-\Psi)$ data. The dashed lines show the scatter for the MANGA and Atlas3D galaxies. Red lines are for the SRs, blue lines for the FRs. The triaxiality of the IllustrisTNG FRs is consistent with the  near-alignment of their kinematic and photometric position angles.}
    \label{fig:misalignment1re}
\end{figure}

\begin{figure}  
    \centering
    \includegraphics[width=\linewidth]{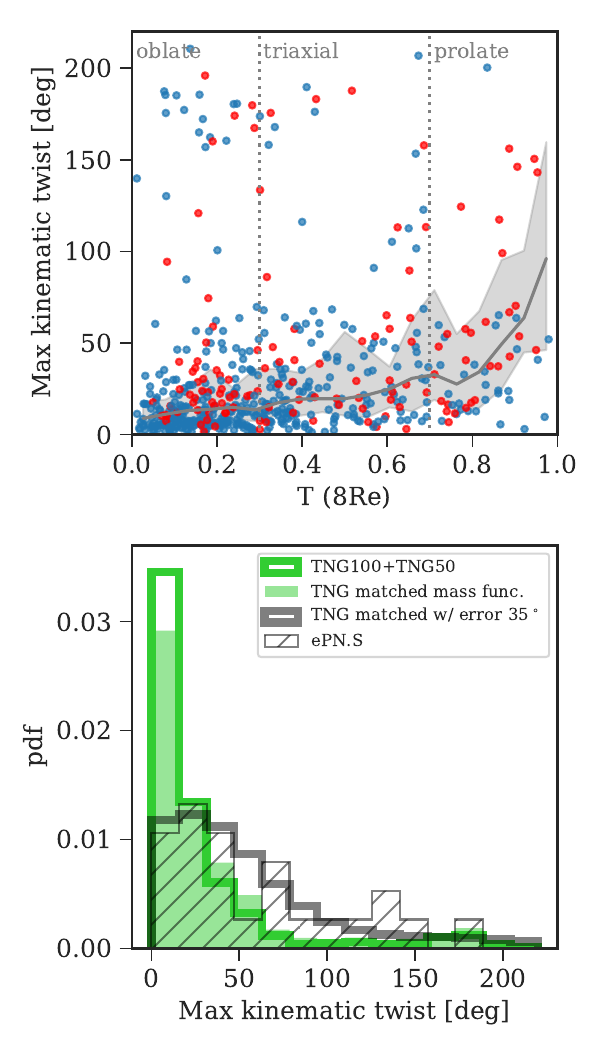}
    \caption{\textbf{Top}: maximum kinematic twist measured between 1 and 8 \re{} as a function of the stellar halo triaxiality $T(8\re{})$. FRs are shown in blue, SRs in red; both TNG50 and TNG100 samples are shown. The gray line and shaded band show the median of the twist distribution as a function of $T(8\re{})$ and the 25th and 75th percentiles. Even though the majority of the TNG galaxies have small kinematic twists, high $T$ galaxies are more like to display a kinematic twist. \textbf{Bottom}: Distribution of the maximum kinematic twist for the TNG50 and TNG100 samples (green step histogram) compared with the ePN.S sample (hatched histogram). The filled green histogram shows the distribution of the TNG galaxies after their mass function is matched to that of the ePN.S sample; the gray histogram is the mass-matched TNG distribution convolved with a measurement error and is consistent with the data.}
    \label{fig:Kintwists}
\end{figure}

Kinematic signatures of galaxy triaxiality can be found from observations of minor axis rotation, kinematic twists, and misalignment of the kinematic position angle $\PAkin{}$ with the photometric $\PAphot{}$ \citep[see e.g.][]{1985MNRAS.212..767B, 1991ApJ...383..112F}. 

Figure \ref{fig:misalignment1re} shows the distribution of misalignments $\Psi = \PAkin{}(1\re{})-\PAphot{}(1\re{})$ as a function of the ellipticity for the random LOS projected TNG galaxies. The solid lines show the unweighted root-mean-square deviation from zero of the data points as a function of the ellipticity for the FR and the SR classes separately, computed by mirroring the data points around zero (i.e. using $[\Psi, -\Psi]$). The dashed lines show the same quantities for observed galaxies in Atlas3D and MANGA \citep{2011MNRAS.414.2923K, 2018MNRAS.477.4711G}. Simulated and observed galaxies show very similar trends with ellipticity, with very flat galaxies being strongly aligned, and kinematic misalignment increasing with rounder shapes. Simulated FRs are found to be much more aligned than SRs, in agreement with observations.

Data points in Fig. \ref{fig:misalignment1re} are color coded according to the intermediate to major axis ratio $p$ at 1\re{}. At these radii the errors on the axis ratios are at most 0.1 for the low mass systems (see App.\ref{appendix:accuracy_shapes_measurement}), so $p(1\re{})$ measurements are generally well defined. We find that many TNG galaxies, both SRs and FRs show a high degree of alignment ($|\Psi|<10^\circ$) even though they are far from being oblate (i.e. with $p<0.9$). This result agrees with the analysis of \cite{2019MNRAS.487.2354B} of Illustris galaxies, among which they found many triaxial and prolate objects with small $\Psi$. Although we can not draw conclusions on the shape distribution of the real galaxies, Fig. \ref{fig:misalignment1re} implies that near-alignment of the kinematic and photometric position angles does not exclude triaxiality for the IllustrisTNG FR galaxies even at $1\re{}$.

Simulated FRs and SRs do show kinematic position angle variations as a function of radius. An example is the object shown in Figs. \ref{fig:photometry_example} and \ref{fig:velfield_example}, a galaxy with central disk embedded in a triaxial stellar halo, which shows a variation in the direction of rotation corresponding to the sudden change in the triaxiality profile around $r\sim5\re{}$. 

The top panel of Fig. \ref{fig:Kintwists} shows the distribution of the maximum kinematic twists, i.e. the maximum variation of $\PAkin{}$, measured between 1 and 8 \re{} for all the TNG galaxies, as a function of the stellar halo triaxiality measured at $8\re{}$, i.e. where the median $T$ profiles are constant with radius (see Fig. \ref{fig:triaxiality_profiles}). 

We find that the large majority of TNG galaxies have small kinematic twists compared to typical measurement errors of $\PAkin{}$ from discrete tracers (the median error for the ePN.S galaxies is $\sim35^\circ$). Aside for a small group of near-oblate galaxies with counter-rotating disks (twist $\sim180^{\circ}$), the main dependence of the amplitude of the twist with the triaxiality parameter is such that galaxies with high $T$ are more likely to have large kinematic twists, as shown by the solid gray line in Fig. \ref{fig:triaxiality_profiles} representing the median twist as a function of $T$. This highlights the importance of kinematics as a fundamental tool to investigate the intrinsic structure of galaxies besides photometry alone (see Sect. \ref{sec:signatures_triaxiality_photometry}). On the other hand Fig.  \ref{fig:triaxiality_profiles} points out that not all the triaxial and prolate galaxies in IllustrisTNG display kinematic twists. This suggests that also in observations a galaxy's triaxiality may not be easily revealed by kinematic misalignments.

IFS kinematics show that the central regions of FR have $\PAkin{}$ and $\PAphot{}$ aligned within $\sim 10$ degrees, while SRs are generally misaligned \citep{2011MNRAS.414.2923K, 2015MNRAS.454.2050F, 2018MNRAS.479.2810E, 2018MNRAS.477.4711G}. This difference in the misalignment distribution was interpreted as signature of a different intrinsic shape distribution for the two classes, with the FRs being consistent with having oblate shapes, and the SRs being moderately triaxial. This interpretation is not supported by the results in Fig. \ref{fig:misalignment1re}.

Even though the central regions of FRs have been found to have $\PAkin{}$ well aligned with $\PAphot{}$, kinematic twists are observed in the halos. By extending the kinematic study at larger radii using planetary nebulae, \cite{2018A&A...618A..94P} found that kinematic twists are relatively frequent in the ePN.S sample of FRs ($\sim40\%$). They concluded that if the central regions of FRs are oblate, kinematic twists would indicate a transition to halos with triaxial shapes. 

In the bottom panel of Fig. \ref{fig:Kintwists} we compare the distribution of kinematic twists in the IllustrisTNG galaxies with the observed distribution in the ePN.S sample. To do that we match the TNG mass function to that of the ePN.S sample by randomly selecting the appropriate fraction of TNG galaxies in mass bins. The filled green histogram in Fig. \ref{fig:Kintwists} shows the distribution of kinematic twists for 100 random realizations of PNS-like samples extracted from the TNG galaxies. We then convolved the resulting distribution with a Gaussian error of $35$ degrees, i.e. the median error of the ePN.S measurements (gray histogram). We find that the simulated galaxies show a similar distribution as the ePN.S galaxies, although there is an indication for a lower fraction of galaxies with large kinematic twists in IllustrisTNG. This might be due to a different sample selection between simulations and ePN.S, with the former potentially containing a larger fraction of disk galaxies even after the matching of the mass functions. Figures \ref{fig:MassFunction} and \ref{fig:ElliptDistr}a in fact show that the ePN.S sample is on average more massive and contains rounder galaxies than TNG100.


\section{Stellar halo angular momentum and shape}
\label{sec:AM_shape}

\begin{figure}  
    \centering

    \includegraphics[width=0.8\linewidth]{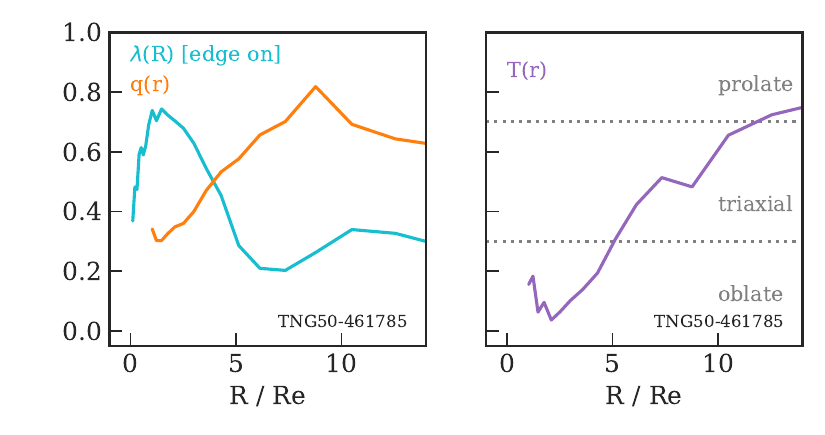} 
    \includegraphics[width=0.95\linewidth]{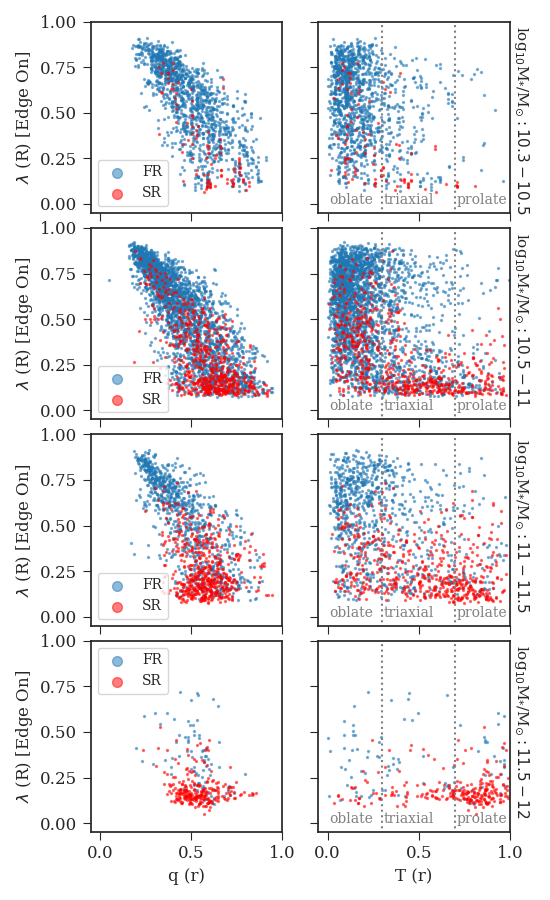}    
    \caption{\textbf{Top}: $\lambda(R)$, axis ratio $q(r)$ and triaxiality $T(r)$ profiles for an example galaxy.
    \textbf{Bottom}: $\lambda$ for the edge-on projection versus minor to major axis (left panels) and triaxiality parameter (right panels) in stellar mass bins. Each data point is a local measurement within a shell of major axis $R = r$: each TNG galaxy is represented by $\sim 6$ data points measured between $3.5-8\re{}$. FRs and SRs are shown with different colors as in the legend. The mass bins are labelled on the right margins. Lower rotational support $\lambda$ in the stellar halos is related to rounder shapes, with a wide range of triaxiality which depends on stellar mass. FRs and SRs exhibit a continuity in stellar halo properties rather than a bimodality.}
    \label{fig:halo_AMandT}
\end{figure}

In Sect. \ref{sec:simulated_VS_observed_rotation_halo} we observed that the rotational support in the stellar halo correlates with the intrinsic shape: stellar halos with high  $V_\mathrm{rot}/\sigma$ are likely near-oblate, while halos with lower $V_\mathrm{rot}/\sigma$ can have larger triaxiality. In this section we connect the variation of rotational support in the stellar halos to variations of their intrinsic shape.

The top panel of Fig. \ref{fig:halo_AMandT} shows the differential $\lambda(R)$ (measured for the edge on projection), the axis ratio $q(r)$, and the triaxiality parameter profile $T(r)$ for an example galaxy. It has a decreasing $\lambda(R)$ profile in the halo, while $q(r)$ and $T(r)$ increase: for this object the decreased rotational support marks a variation in the intrinsic shape of the galaxy from relatively flat and near-oblate at $R\sim2\re{}$, where the rotation is highest, to triaxial spheroidal in the outskirts. In observations the drop in rotation of the ePN.S FRs is often related to a decrease in ellipticity. This led to the idea that central fast rotating regions of FRs are embedded in a more dispersion dominated spheroidal stellar halo. 

We verify this conclusion in Fig. \ref{fig:halo_AMandT}. Here the outskirts of galaxies are divided into 6 ellipsoidal shells of major axis $r$ between $3.5$ to $8\re{}$. This radial range is motivated by the requirement that $r$ is large enough so that the errors on the $T$ parameter are small for galaxies of all masses, but also such that most of the galaxies (96\%) have enough particles at large radii (8\re{}) in order to measure their intrinsic shapes. In each ellipsoidal shell of major axis $r$ and width $\Delta r$ we measure the axis ratio $q$ and the triaxiality parameter. Then we measure the edge on projected $\lambda(R)$ in an elliptical shell of major axis $R=r$ and width $\Delta R = \Delta r$. 
Figure \ref{fig:halo_AMandT} shows the halo edge-on projected $\lambda(R)$ for all shells and all TNG galaxies in four galaxy stellar mass bins, as a function of the minor to major axis ratio $q(r)$ and of the triaxiality parameter $T(r)$.

We find indeed a relation between $\lambda$ and shape. High rotational support is related to flattened (i.e. low $q$) near-oblate (i.e. low $T$) shapes. Where $\lambda$ decreases $q$ grows towards more spheroidal shapes, although the scatter in possible stellar halo shapes is large ($\sigma_q \sim 0.15$). Stellar halos of both FRs and SRs with high triaxiality generally have low $\lambda$, but lower $T$ is compatible with all values of $\lambda$. This means that the decrease in the rotational support of the TNG FRs follows a change in the structure of the galaxies, which become more spheroidal in the outskirts. These outer spheroidal components can span all values of triaxiality.
By comparison to the FRs, the SRs show smaller variations in both intrinsic shapes (Fig. \ref{fig:triaxiality_profiles}) and $\lambda$.

Figure \ref{fig:halo_AMandT} confirms the dependence of the stellar halo intrinsic shape and rotational support on stellar mass already described in Sects.   \ref{sec:halo_intrinsic_shape_distr}, \ref{sec:triaxiality_profiles}, and \ref{sec:lambda_profiles}. Lower mass galaxies host more rotationally supported stellar halos with near-oblate shapes. At progressively higher masses the fraction of dispersion dominated stellar halos increases as well as the fraction of halos with high triaxiality. 

It is interesting to note that there is no clear separation in Figure \ref{fig:halo_AMandT} between the stellar halo properties of the FRs and SRs, but rather a continuity of properties among the two classes, with the low $T$-high $\lambda$ extreme dominated by the FRs, the high $T$-low $\lambda$ limit by the SRs, and the relative importance of the two populations gradually changing with stellar mass. This result implies that there is no qualitative difference between the structure of the galaxies at large radii despite the bimodality of the FR/SR classification of the centers. The IllustrisTNG galaxies agree with the ePN.S observations in that FRs and SRs tend be more similar at large radii, especially at intermediate to high masses.

The similarity of the overall shapes of the $\lambda(R)$ (or $V_\mathrm{rot}/\sigma(R)$) profiles and of the $\varepsilon(R)$ profiles between ePN.S and TNG galaxies described in in Sects.  \ref{sec:ellipticity_profiles} and \ref{sec:simulated_VS_observed_rotation_halo} suggests that intrinsic shape variations similar to those found in the TNG galaxies also exist in real galaxies.


\section{Summary and conclusions}\label{sec:summary_and_conclusions} 

In this paper we have analysed the kinematic and photometric properties of $\sim1200$ early type galaxies (ETGs) from the IllustrisTNG cosmological simulations TNG100 and TNG50, with a focus on their stellar halos. The sample of simulated ETGs was selected in stellar mass and in color (Fig. \ref{fig:SelectionSample}) and in the $\lambda_e-\varepsilon$ diagram (Fig. \ref{fig:lambda_ell_center}). There we excluded simulated objects that do not match the observed properties in the central regions of fast rotators (FRs) and slow rotators (SRs), and that appear as highly centrally elongated red galaxies. We verified that this does not affect our results on the stellar halo properties of the simulated galaxies. The resulting ETG sample has mass-, size-, and central kinematics distributions consistent with observations.

For the selected sample we determined mean velocity fields, kinematic, and photometric profiles, and studied the intrinsic shapes of the simulated galaxies from the central regions into their outskirts, up to $15\re{}$. The purpose of the paper is to study the kinematic properties of stellar halos and connect them to variations in the structural properties of their galaxies from the central regions to the halos. Our conclusions are as follows:

\vspace{6pt}
1) The differential $\lambda$ profiles and the triaxiality profiles (Figs. \ref{fig:triaxiality_profiles} and \ref{fig:lambda_profiles}) successfully reproduce the diversity of kinematic and photometric properties of stellar halos observed in the ePN.S survey. 
\begin{itemize}
    \item We find that simulated FRs divide almost evenly among a third with flat $\lambda$  profiles, a  third with gently decreasing profiles, and another third with very low rotation in the halo. Half of the SRs do not show any rotation in the halo, while the other half has increased rotational support at large radii. 
    
    \item FRs generally tend to show increased triaxiality with radius, although the majority (partially driven by the numerous low mass fast rotators) have stellar halos consistent with oblate shapes.
    
    \item Both halo triaxiality and rotational support are found to depend on stellar mass, with higher mass galaxies being more triaxial and more dispersion dominated at large radii.
    \end{itemize} 

2) Halo intrinsic shape and rotational support are strongly related (Fig. \ref{fig:halo_AMandT}):
\begin{itemize}
    \item High $\lambda$ is related to flattened oblate shapes. 
    \item Where $\lambda$ decreases with radius, galaxies tend to become rounder, but with a wide range of triaxiality. 
    
\end{itemize} 

\vspace{6pt}
3) The FR class in TNG shows the largest variety in stellar halo properties and the largest variations with radius in both intrinsic shapes and rotational support. Among these galaxies we can find rotationally supported stellar halos with flattened oblate shapes, as well as FRs that have central rotating disk-like structures embedded in more spheroidal components. For a subset of these the stellar halos can reach high triaxiality values. SRs, by comparison, display milder changes in structure with radius.

\vspace{6pt}
4) The TNG FRs exhibit shallower $V_\mathrm{rot}/\sigma(R)$ profiles than the Atlas3D and ePN.S galaxies (Fig. \ref{fig:comparisonA3D}). Both TNG50 and TNG100 FRs tend to reach a peak in rotation at a median radius of $\sim3\re{}$, compared to $\sim1-1.3\re{}$ for the Atlas3D and ePN.S galaxies. This result implies a more extended distribution of the angular momentum with radius in the TNG galaxies than observed.

\vspace{6pt}
5) However, even though the $V_\mathrm{rot}/\sigma(R)$ profiles do not agree quantitatively between simulated and observed ETGs galaxies, the simulations do reproduce the observed kinematic transitions between central regions and outskirts. The similarity between the scaled shapes of the $V_\mathrm{rot}/\sigma(R)$, and the $\varepsilon(R)$ profiles between ePN.S and TNG galaxies (Figs. \ref{fig:ElliptDistr}b and \ref{fig:Vsigma}) suggests that also the observed variations in the kinematics between central regions and halos trace changes in the intrinsic structure of the galaxies. 

\vspace{6pt}
6) We find that most of the triaxial TNG galaxies display modest photometric twists that only weakly depend on triaxiality (Fig. \ref{fig:Phot_twist}). For these galaxies kinematic twists are larger (Fig. \ref{fig:Kintwists}), but in many cases they are not large enough to be measured by currently available data.
    
\vspace{6pt}
7) By comparing the distributions of the minor-to-major axis ratios $q$, triaxiality parameters $T$, and angular momentum parameters $\lambda$ (Fig. \ref{fig:halo_AMandT}),  we find that lower rotational support in the stellar halos is related to rounder shapes, with a wide range of triaxiality which depends on stellar mass.  In this there is no qualitative difference between the FRs and SRs. Rather, despite the bimodality of the central regions, the two classes show a continuity of halo properties with the FRs dominating the low $T$-high $\lambda$ end of the distribution and the SRs dominating in the high $T$-low $\lambda$ extreme. The relative weight of the different sides of the distribution gradually changes with stellar mass. This is in agreement with ePN.S observations of ETG halos.

\vspace{6pt}
In a companion paper we will investigate the dependence of the stellar halo parameters on the accretion history of galaxies, and explore the relation between stellar and dark matter halo properties.

\begin{acknowledgements}
  \\
  We thank the anonymous referee for helpful comments that improved the clarity of the paper. C.P. is extremely grateful to F. Hofmann for his support.
  This research has made
  use of the NASA/IPAC Extragalactic
  Database (NED).
\end{acknowledgements}

\bibliographystyle{aa}
\bibliography{auto}


 \begin{appendix}

\section{Color transformation equations} \label{sec:color_relations}

\begin{figure}[ht]
    \centering
    \includegraphics[width=0.49\linewidth]{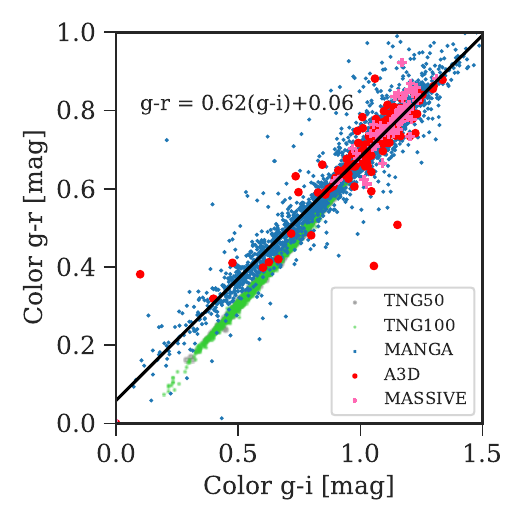}
    \includegraphics[width=0.49\linewidth]{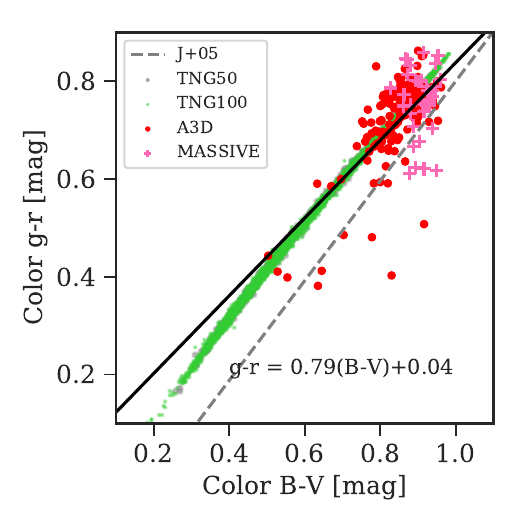}
    \caption{Relation between colors for the observed galaxies, as shown in the legend. The simulated galaxies with $M_{*}\geq10^{10}M_{\odot}$ are also shown but without any modeling for dust attenuation, which probably explains why they tend to follow a different color relation than the observations. The solid lines show the linear fit to the observations, while the dashed line shows the color transformation equation from \cite{2005AJ....130..873J}.}
    \label{fig:color_relations}
\end{figure}

In Sect.  \ref{sec:SelectionSample} we selected a sample of ETGs from TNG50 and TNG100 by using a cut in $g-r$ color and stellar mass, and compare with observations. Most of the observed galaxies have $g-r$ colors in the NSA catalog. For a small fraction of the observations, only $B-V$ or $g-i$ colors are available. \cite{2005AJ....130..873J} published transformation equations between the SDSS and UBVRcIc magnitudes valid for stars and quasars, which provide reasonable results for galaxies. We note, however, that these transformation equations do not readily apply to the observed galaxies. In Fig.  \ref{fig:color_relations} we show the relation between colors in galaxies that have measured both $g-r$ and $B-V$, and $g-r$ and $g-i$. $g-r$ and $g-i$ colors are from the NSA catalog, $B-V$ colors are from Hyperleda. We find that the observed ETGs follow the relations:
\begin{equation}
    \begin{split}
        g-r & = 0.79 (g-i) - 0.04\\
        g-r & = 0.62 (B-V) - 0.06.
    \end{split}
\end{equation}
Hence we use these derived transformation equations to transform the measured $B-V$ or $g-i$ colors in $g-r$.

\section{Accuracy on the measured intrinsic shapes and angular momentum parameter $\lambda_e$}
\label{appendix:accuracy_shapes_measurement}

Physical properties measured in simulated galaxies are affected by resolution effects coming from the discrete particle nature of these systems. In a collisionless dark matter-only (DMO) simulation the resolution of the gravitational potential depends on the softening length and the particle mass resolution (i.e. the number of particles), which particularly affect measurements in the central regions of the simulated systems \citep[e.g.][]{2003MNRAS.338...14P}. In a full physics simulation instead, like TNG100 and TNG50, the resolution has additional effects on the baryon physics, which require the models to be calibrated on observations \citep{2015MNRAS.446..521S, 2018MNRAS.473.4077P}. In this section we are not concerned with the latter effects, which also have an impact on the galaxy properties. Here we aim at quantifying the effects from the resolution of the gravitational potential on shape and spin measurements at $1\re{}$ and beyond.

\cite{2019MNRAS.484..476C} analyzed the convergence of intrinsic shape profiles in the Illustris-DMO simulation, with the procedure recommended by \citet{2011ApJS..197...30Z}, also used in this paper and outlined in Sect. \ref{sec:measuring_IntrinsicShape}. From their Figure 1 we find that the shape profiles are converged within 0.1 in both $p$ and $q$ already at $r\sim 2\rsoft{}$. Since TNG100 has similar particle resolution as Illustris-DMO (and smaller $\rsoft{}$), we can apply a similar convergence criterion in TNG100. We verified that the two simulations have comparable particle numbers at $r\sim 9\rsoft{}$, i.e. where full convergence is achieved according to the prescription of \cite{2019MNRAS.484..476C}. At smaller radii the full physics TNG100 simulation has more particles than the DMO simulation since the baryons are subjected to dissipation; hence we expect similar, if not better, convergence in TNG100. 

Twice the softening radius in both TNG50 and TNG100 corresponds to a radius $\lesssim \re{}$ for all the selected galaxies (see Fig. \ref{fig:MassSize}, and note that in TNG100 we excluded the galaxies with $\re{}<2\rsoft{}$ from the sample). 
Since $\re{}$ increases with stellar mass, the more massive systems are better resolved. For example, in TNG100 the median effective radius at $M{*}=10^{11}\MSUN{}$ is $1\re{}\sim 7.8\rsoft{}$, where the resolution effects on both $p$ and $q$ are $\lesssim 0.02$.
Thus the absolute error on the axes ratios $p$ and $q$ measured at $r=1\re{}$ is mass dependent, and it is at most 0.1 for the low mass systems.
At $r\gtrsim 9\rsoft{}$, which corresponds to $r\gtrsim3.5\re{}$ at the low mass end, and to $1.16\re{}$ for $M_{*}=10^{11}\MSUN{}$, the softening effects are negligible compared to the errors coming from particle statistics. For a minimum required number of 1000 particles per ellipsoidal shell we found that these errors are generally $\lesssim0.02$ on both $p$ and $q$.
Hence at $r\gtrsim 9\rsoft{}$ the uncertainty on the axes ratios is of the order of 0.02.

Throughout the paper we quantify the intrinsic shapes of galaxies also with the triaxiality parameter. Because of its definition (Eq. \eqref{eq:triaxiality}), the error $\Delta T$ is shape dependent, as well as mass dependent as discussed above. In this work we consider reliable $T$ measurements performed at $r\geq9\rsoft{}$, where the uncertainties on the axes ratios are $\sim0.02$, corresponding to $\Delta T \sim 0.2$ for typical FRs axis ratios (i.e. $p=0.9$ and $q=0.5$). At smaller radii, where $\Delta T$ grows large for the low mass galaxies, we quantify the intrinsic shapes using only the better determined $p$ and $q$ (e.g. in Fig.  \ref{fig:misalignment1re}).

The angular momentum parameter $\lambda_e$ is evaluated by integrating over all the particles within $R\leq1\re{}$, hence, by definition, it is derived more reliably than the flattening. This is apparent in the convergence study of \cite{2017MNRAS.464.3850L} on the EAGLE simulations, which have particle mass resolution and gravitational softening length very similar to TNG100. The study shows that the angular momentum within $1\re{}$ is well converged already at stellar masses above $M_{*}=10^{9.5}M_{\odot}$, i.e. within galaxies resolved by about 2000 particles. The selected galaxies in the TNG100 sample are resolved with more than $2\times10^4$ particles, hence the measured $\lambda_e$ is independent of the softening and particle mass resolution. 

In conclusion we find that both angular momentum and shapes are not (or only marginally in case of the shape) affected by the resolution of the gravitational potential and by the particle number at $r\gtrsim1\re{}$, and they are well-determined at larger radii.

\section{Elongated galaxies in IllustrisTNG} 
\label{appendix:elongated_galaxies}

In Sect.  \ref{sec:SelectionSampleII} we further restrict the sample selection by excluding an excess of centrally elongated galaxies not present in the observed samples.  The inconsistency is revealed in the $\lambda_e-\varepsilon(1\re{})$ diagram of Fig. \ref{fig:lambda_ell_center}, in which the centrally elongated galaxies distribute in a region at high ellipticity where there are few observed counterparts. In App.\ref{appendix:accuracy_shapes_measurement} we showed that the resolution effects at $1\re{}$ on the intrinsic shapes are at most of the order of 0.1: these are not enough to explain the differences with observations in the $\lambda_e$-ellipticity diagram, nor the extreme values measured for the flattening $p(1\re{})$.

\begin{figure}  
    \centering
    \includegraphics[width=\linewidth]{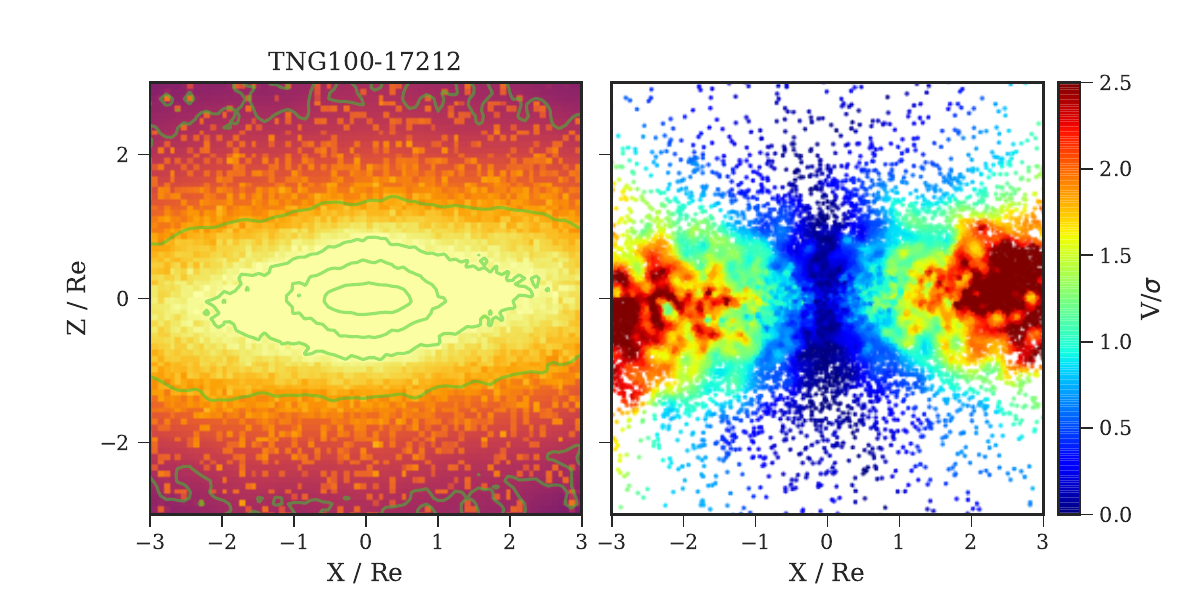}
    \includegraphics[width=\linewidth]{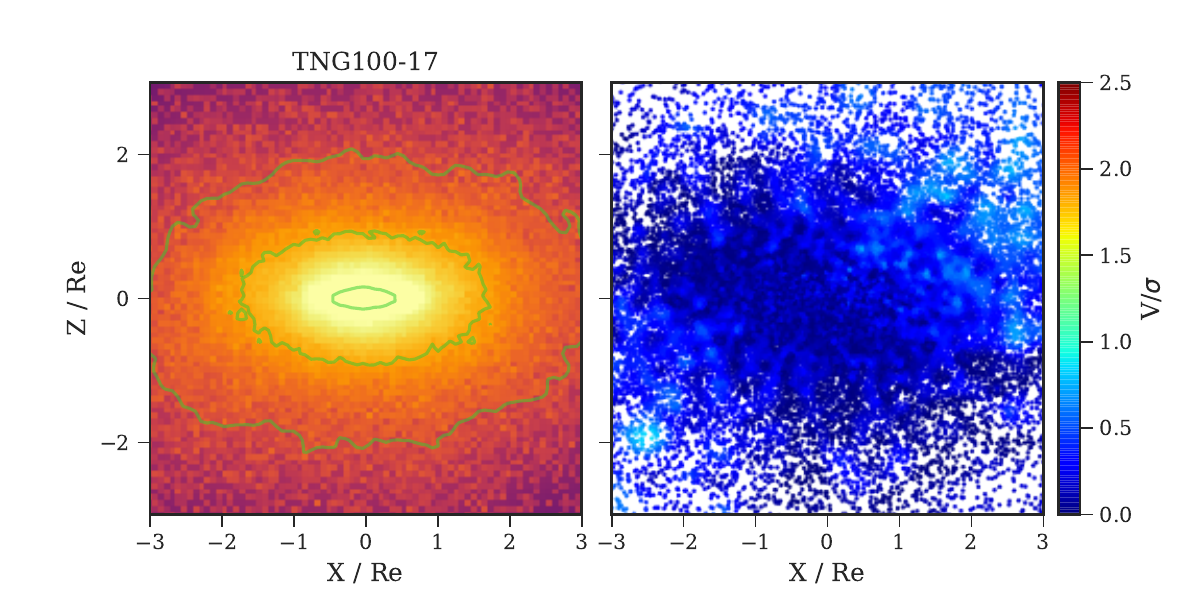}
    \caption{Stellar mass distribution (left) and $V/\sigma$ fields (right) of the central 3 Re for two examples of the centrally elongated galaxies projected edge-on. In the top panel the elongated central component ($p(1\re{})=0.42$, $q(1\re{})=0.33$, and $\lambda_e=0.54$)
    is embedded in a disk structure ($q(5\re{})=0.33$). In the bottom panel the bar-like structure ($p(1\re{})=0.39$, $q(1\re{})=0.30$, and $\lambda_e=0.10$) is embedded in a spheroid ($q(5\re{})=0.61$).} 
    \label{fig:velocity_fields_elongated}
\end{figure}

\begin{figure}  
    \centering
    \includegraphics[width=\linewidth]{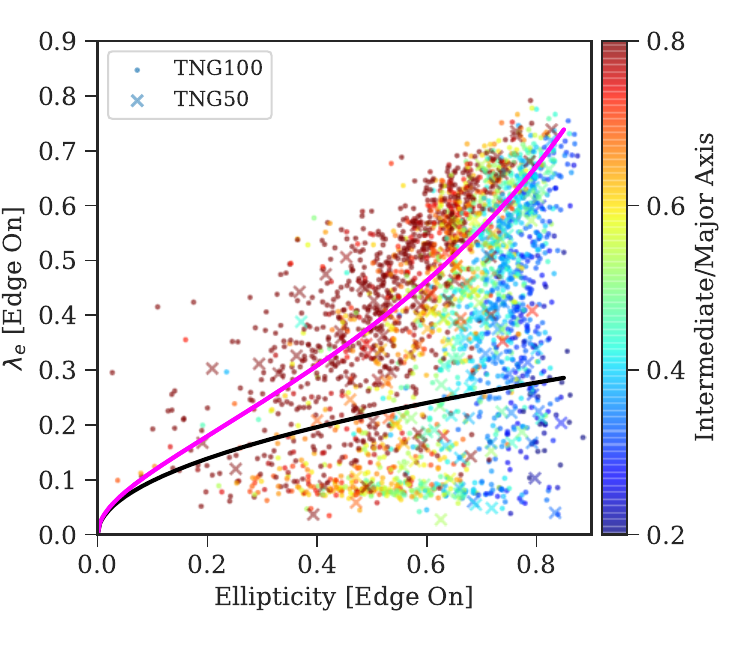}
    \caption{$\lambda_e$-ellipticity (1\re{}) diagram for the TNG galaxies selected in color and stellar mass as in Sect. \ref{sec:SelectionSample}, and - differently from Fig.~\ref{fig:lambda_ell_center} - projected edge-on. The symbols are colored according to the intermediate to major axis ratio measured at 1$\re{}$. The black and magenta lines are as in Fig. \ref{fig:lambda_ell_center}.}
    \label{fig:lambda_ell_edgeon}
\end{figure}

Figure \ref{fig:velocity_fields_elongated} shows example velocity fields for two of these centrally elongated objects (i.e. with $p(1\re{})<0.6$) projected edge-on. The colors show the line-of-sight mean velocity divided by the velocity dispersion at the position of each particle. The two galaxies shown have similar flattening in the central regions $p(1\re{})\sim0.4$ and $q(1\re{}\sim0.3)$, but in one case (top panel) the galaxy contains an extended disk, in the other the galaxy is spheroidal in the outskirts (bottom panel). 
Most of the selected systems with $p(1\re{})<0.6$ have velocity fields with rotation around the intrinsic minor axis (as the example in Fig.\ref{fig:velocity_fields_elongated}). However, Fig.\ref{fig:lambda_ell_edgeon} shows that these galaxies (in blue to yellow colors) can exhibit different degrees of edge-on rotation at 1 $\re{}$ and some of them do not rotate at all (see also bottom panel of Fig. \ref{fig:velocity_fields_elongated}), which is at odds with what is typically seen in barred galaxies \citep[e.g.][]{2019A&A...632A..59F}.

  \begin{figure}  
    \centering
    \includegraphics[width=\linewidth]{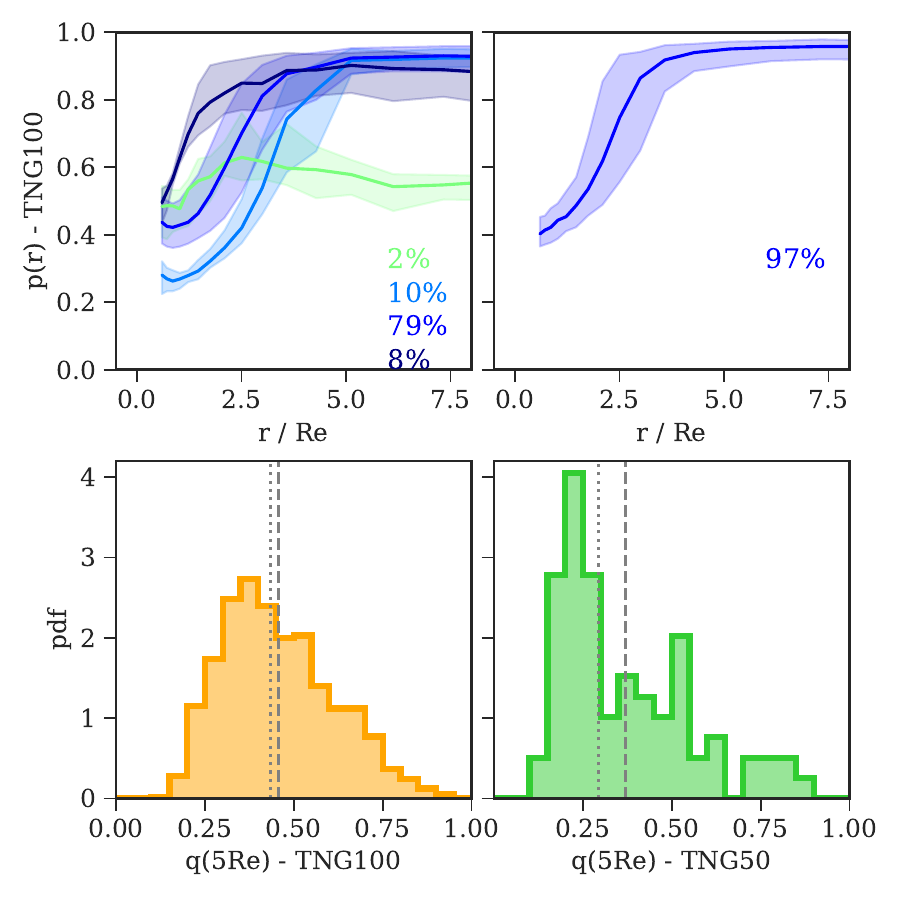}
    \caption{Top: Median $p(r)$ profiles for the galaxies with $p(1Re)< 0.6$ in TNG100 (left) and TNG50 (right). Only bins with more than 10 galaxies are shown. The percentages indicate the fraction of centrally elongated galaxies that populate each median profile. Bottom: Distribution of the axis ratio $q(5\re{})$ in the centrally elongated galaxies of TNG100 (left) and TNG50 (right). Dashed and dotted vertical lines show the mean and the median of the distributions, respectively. The galaxies with $p(1Re)< 0.6$ have elongated shapes up to $\sim 3-4$ Re; outside this regions ($r\sim 5\re{}$) they are mostly near-oblate ($p\sim0.9$) with median flattening $q\sim0.3-0.4$, depending on the simulation.} 
    \label{fig:intrinsic_shape_elongated}
\end{figure}

\begin{figure}  
    \centering
    \includegraphics[width=\linewidth]{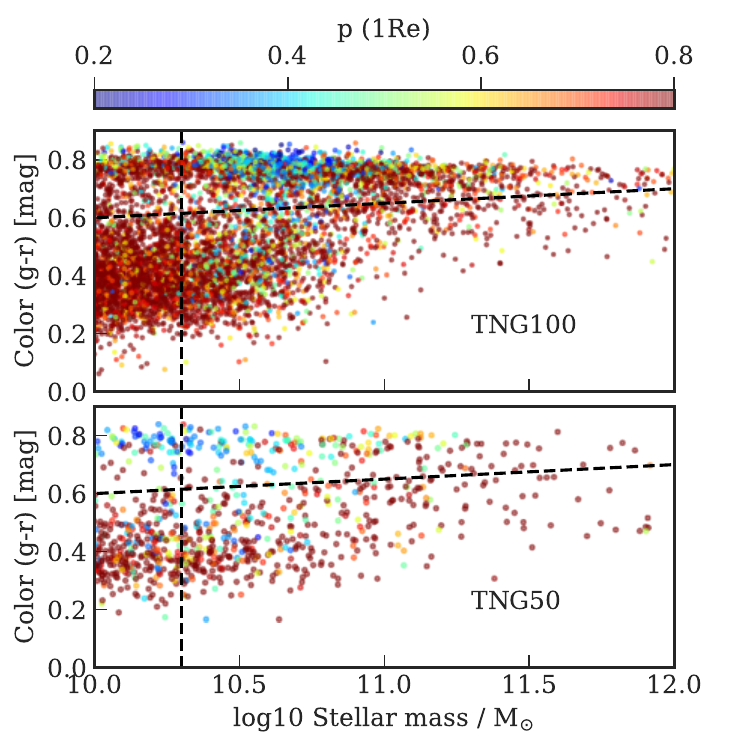}
    \caption{Color (g-r) as a function of the stellar mass. The top panel shows TNG100, the bottom panel TNG50. The dashed lines illustrate the selection of the sample of ETGs with masses $10^{10.3}\leq M_{*} / M_{\odot}\leq 10^{12}$, and red color (Sect. \ref{sec:SelectionSample}). The colors indicate the axis ratios $p(1\re{})$.}
    \label{fig:ColorMass_elongated}
\end{figure}

The top panels of Fig. \ref{fig:intrinsic_shape_elongated} show the median $p(r)$ profiles for the centrally elongated galaxies. The profiles are built by binning together the galaxies according to the values of $p(1\re{})$ and $p(6\re{})$. The percentages next to each profiles gives the fraction of centrally elongated TNG100 or TNG50 galaxies that populate the median profile. The elongated regions can extend up to a few Re, typically $\sim3\re{}$. Outside these radii, galaxies are near-oblate, with median $p(5\re{})\sim 0.9$, and a large scatter on the flattening $q(5\re{})$ (bottom panels in 
Fig. \ref{fig:intrinsic_shape_elongated}). The TNG100 galaxies have rather spheroidal shapes with $q(5\re{})\sim0.45$; the TNG50 galaxies tend to be flatter with $q(5\re{})\sim0.3$, and a peak in the distribution at $\sim0.2$. 

Finally Fig. \ref{fig:ColorMass_elongated} shows the color-stellar mass diagram for the galaxies in TNG100 and TNG50 separately. The axis ratio $p(1\re{})$ is shown by the color of the data points. 
The centrally elongated systems occur prominently among the redder simulated galaxies, and in specific stellar mass ranges. In TNG100 they are numerous in the interval $10^{10.4}\MSUN{}\lesssim M_{*}\lesssim10^{10.9}\MSUN{}$, while in TNG50 they have  $10^{10.0}\MSUN{}\lesssim M_{*}\lesssim10^{10.5}\MSUN{}$.

The fact that these centrally elongated galaxies are preferentially produced in a particular mass range, and that most of them are old, red systems, points towards these objects being a class of galaxies that are produced by the simulation but are not present in nature, probably related to the way the simulated galaxies accrete gas, form stars, and quench during a rapid dynamical evolution in this specific mass range. The difference between TNG100 and TNG50 in the stellar mass range where the centrally elongated systems are produced could be explained by the higher star formation efficiency in the higher resolution simulation \citep{2018MNRAS.473.4077P}. This hypothesis is strengthened by the absence of a similar concentration of centrally elongated systems among the red galaxies of intermediate masses in the Illustris simulation: hence the presence of this population of galaxies is due to the galaxy formation model. We note that in TNG100 this mass range approximately coincides with the knee in the stellar mass-halo mass relation \citep[e.g.][and reference therein]{2013ApJ...770...57B}, where accretion could be particularly efficient. 

As Fig. \ref{fig:intrinsic_shape_elongated} showed, these centrally elongated systems are embedded in an near-oblate component with various flattening $q$, and $q$ tends to be smaller in TNG50. This difference in flattening is likely due to the better resolution of TNG50, which allows to resolve thinner disks \citep[][their appendix B and C]{2019MNRAS.490.3196P}. From App. \ref{appendix:accuracy_shapes_measurement} above the error on $q$ due to the spatial resolution is of the order of 0.1 at $\sim 1\re{}$ for the lowest mass systems, so that the measured thickness of a thin disk with radius $\sim5\re{}$ and $q\sim0.2$ may be subject to a similar uncertainty. If the centrally elongated galaxy components in TNG100 and TNG50 actually form within disks, then they could be the result of a bar instability, such that at the particular stellar mass range where the bars are produced, too much cold gas forms stars too quickly, while building massive bar-unstable disks. Then the feedback from the intense star formation would sweep away the remaining gas, quench the star formation, and lead to the formation of preferentially red bar-like inner components. For some of these objects the time-scale for dynamical friction against the surrounding stars and dark matter might be short enough to slow down their rotation, generating the wide range of $\lambda_e$ values seen in Fig. \ref{fig:lambda_ell_edgeon}.  

This suggests that these centrally elongated galaxies may be systems that were in the process of forming a disk, whose evolution was interrupted or derailed by rapid dynamical instability, star formation, and feedback in the simulations, in the particular mass range in which they occur. Fig. \ref{fig:bias_in_halo_props} in the main text demonstrates that the distributions of angular momentum and triaxiality in the surrounding stellar halos are not affected by the central elongation $p(1\re{})$ of the simulated galaxies. Thus these over-elongated systems can simply be excluded from the sample selected for our main analysis.

  \end{appendix}

\end{document}